\documentclass[aps,prx,reprint,superscriptaddress,longbibliography]{revtex4-1}
\usepackage{graphicx,epsfig}
\usepackage{amsmath,amssymb,bm}
\usepackage{color}
\usepackage[usenames,dvipsnames]{xcolor}
\usepackage[colorlinks=true,linkcolor=Maroon,citecolor=OliveGreen,urlcolor=Blue,linktoc=page]{hyperref}
\usepackage{placeins}
\usepackage{pdfpages}
\usepackage{pgffor}

\makeatletter
\AtBeginDocument{\let\LS@rot\@undefined}
\makeatother

\begin{document}

\title{
\textcolor{black}{Unusual dynamic charge correlations in simple-tetragonal HgBa$_{2}$CuO$_{4+\delta}$ }
}

\author{B.~Yu}
  \affiliation{School of Physics and Astronomy, University of Minnesota, Minneapolis, Minnesota 55455, USA}
\author{W.~Tabis}
  \affiliation{School of Physics and Astronomy, University of Minnesota, Minneapolis, Minnesota 55455, USA}
  \affiliation{Institute of Solid State Physics, TU Wien, 1040 Vienna, Austria}
  \affiliation{AGH University of Science and Technology, Faculty of Physics and Applied Computer Science, 30-059 Krakow, Poland}
\author{I.~Bialo}
  \affiliation{Institute of Solid State Physics, TU Wien, 1040 Vienna, Austria}
  \affiliation{AGH University of Science and Technology, Faculty of Physics and Applied Computer Science, 30-059 Krakow, Poland}
\author{F.~Yakhou}
  \affiliation{European Synchrotron Radiation Facility, 71 Avenue de Martyrs, CS40220, F-38043 Grenoble Cedex 9, France}
\author{N.~B.~Brookes}
  \affiliation{European Synchrotron Radiation Facility, 71 Avenue de Martyrs, CS40220, F-38043 Grenoble Cedex 9, France}
\author{Z.~Anderson}
  \affiliation{School of Physics and Astronomy, University of Minnesota, Minneapolis, Minnesota 55455, USA}
\author{Y.~Tang}
  \affiliation{School of Physics and Astronomy, University of Minnesota, Minneapolis, Minnesota 55455, USA}
\author{G.~Yu}
 \affiliation{School of Physics and Astronomy, University of Minnesota, Minneapolis, Minnesota 55455, USA}
\author{M.~Greven}
 \affiliation{School of Physics and Astronomy, University of Minnesota, Minneapolis, Minnesota 55455, USA}

\date{\today}

\begin{abstract}
The charge-density-wave (CDW) instability in the underdoped, pseudogap part of the cuprate phase diagram has been a major recent research focus, yet measurements of dynamic, energy-resolved charge correlations are still in their infancy. 
\textcolor{black}{Such information is crucial in order to help discern the connection between CDW and pseudogap phenomena, and to understand the extent to which charge correlations in general shape the phase diagram}.
We report a 
resonant inelastic X-ray scattering study of the underdoped cuprate superconductor HgBa$_{2}$CuO$_{4+\delta}$ ($T_c = 70$ K).  At 250 K,
above the \textcolor{black}{previously established temperature $T_\mathrm{CDW} \approx 200$ K that signifies the onset of quasistatic short-range CDW order}, we observe significant dynamic \textcolor{black}{charge} correlations that are \textcolor{black}{broadly peaked} at about 40 meV \textcolor{black}{and centered at the two-dimensional wavevector $\boldsymbol{q}_\mathrm{CDW}$.} This energy scale is comparable to both the superconducting gap and the low-energy pseudogap. 
\textcolor{black}{At 70 K,
we observe a quasistatic CDW peak at $\boldsymbol{q}_\mathrm{CDW}$}, but the dynamic correlations around 40 meV remain virtually unchanged, 
\textcolor{black}{and we identify a new feature:}
dynamic correlations \textcolor{black}{well above the optic phonon range that are broadly peaked in the 150-200 meV range}. A similar \textcolor{black}{energy} scale was previously identified in other experiments as a high-energy pseudogap. 
\textcolor{black}{The observation} of three distinct features in the charge response is highly unusual for a CDW system 
and suggests that charge order in the cuprates is more complex than previously thought. 
\textcolor{black}{We demonstrate that the energy-integrated signal at $\boldsymbol{q}_\mathrm{CDW}$ is fully consistent with prior resonant X-ray scattering work for HgBa$_{2}$CuO$_{4+\delta}$, and that other single-layer cuprates exhibit approximately the same relative strength of high- to low-temperature charge signal. This finding points to the universal existence of significant dynamic charge correlations in the cuprates.} \textcolor{black}{Intriguingly, the two energy scales identified here are also comparable to those of the superconducting glue function extracted from other spectroscopic techniques, consistent with a dual charge and magnetic nature of the pairing glue.}
We further \textcolor{black}{determine} the paramagnon dispersion along [1,0], across $\boldsymbol{q}_\mathrm{CDW}$, \textcolor{black}{and find it to be} consistent with magnetic excitations measured by inelastic neutron scattering. Unlike for some other cuprates, our result points to the absence of a discernible coupling between charge and magnetic excitations. 
\end{abstract}
\maketitle

\section{Introduction}
The high-$T_{c}$ cuprates are doped charge-transfer insulators with lamellar structures that feature the quintessential CuO$_2$ plane \cite{Keimer2015}. 
At moderate and intermediate hole doping, these complex oxides exhibit a partial depletion of the density of states at the Fermi-level (the pseudogap, PG). 
The PG state is characterized by myriad ordering tendencies, including CDW order with a modulation direction along the planar Cu-O bond and a periodicity of 3-4 lattice units
\cite{Tranquada1995,Hoffman2002,Howald2003,WuNature2011,Ghiringhelli2012,Chang2012,Comin2014,Tabis2014,Croft2014,Hashimoto2014,Wu2015,daSilvaNeto2014,daSilvaNeto2015,GerberScience2015,Chang2016,Peng2016,daSilvaNeto2016,Tabis2017,Arpaia2019}. 
Whereas CDW order competes with superconductivity \cite{Chang2012,Chang2016,GerberScience2015}, dynamic CDW correlations have long been argued to play a \textcolor{black}{pivotal} role in shaping the phase diagram (Fig. 1a)
\cite{Castellani1997,Kivelson2003,Caprara2017}. 
However, key questions remain unresolved. The mechanism of CDW formation could be related to Fermi-surface nesting (i.e., a reciprocal-space mechanism) \cite{Hanaguri2004,Shen2005,Wise2008}, or to strong electronic correlations that lead to charge separation (i.e., a real-space mechanism) \cite{Tranquada1995,Abbamonte2005}. 
Furthermore, the relation between CDW correlations and the PG is far from understood, with suggestions that either one is the underlying phenomenon \cite{Castellani1997,Caprara2017,Pelc2019a,Arpaia2019,Hamidian2016}. 
The short correlation lengths indicate that disorder might play an important role \cite{Caplan2015}, but it is still \textcolor{black}{debated} how the CDW in the cuprates becomes static on cooling. 
\textcolor{black}{Finally, whereas a prominent early theoretical proposal argued in favor of instantaneous pairing \cite{Anderson2007}, ultrafast optical measurements indicate that the interaction in the cuprates is most likely retarded  \cite{dalConte2012,dalConte2015}. Although the relevant bosonic excitations are often argued to be of magnetic origin \cite{Abanov2003,Scalapino2012}, there exists indirect evidence that charge correlations play an important role \cite{Fanfarillo2016,Pelc2019a}.}

CDW correlations have been detected in numerous cuprates, primarily via resonant X-ray scattering (RXS) 
\cite{Ghiringhelli2012,Chang2012,Comin2014,Tabis2014,Croft2014,Hashimoto2014,daSilvaNeto2014,daSilvaNeto2015,BlancoCanosaPRB2014,GerberScience2015,Wu2015,Chang2016,Peng2016,daSilvaNeto2016,Chaix2017,Tabis2017}, including HgBa$_{2}$CuO$_{4+\delta}$ (Hg1201) \cite{Tabis2014,Tabis2017}.  
In RXS, the incident photon energy is tuned to the Cu $L_3$ edge to 
\textcolor{black}{discern modulations of the valence electron density in the quintessential CuO$_2$ layers.} The scattered photons can be directly measured in energy-integrated mode \textcolor{black}{(conventional RXS)}, or additionally analyzed by a spectrometer in energy-resolved inelastic mode (RIXS). \textcolor{black}{Conventional RXS} has the benefit of relatively short counting times and enabled the efficient exploration of the doping and temperature dependence of CDW correlations. In principle, the charge dynamics can be measured via RIXS, yet until recently, the best available energy resolution was well above 100 meV. 

We report a RIXS study of the charge dynamics of underdoped Hg1201 ($T_c = 70$ K; see Fig. 1a) across $\boldsymbol{q}_{CDW}$ with energy resolution of 60 meV. \textcolor{black}{Although this energy resolution is still an order of magnitude larger than that of, e.g., THz experiments \cite{Torchinsky2013,Hinton2013}, it is very high for RIXS, and we are able to obtain qualitatively new energy- and momentum-resolved information.}  
Hg1201 is a single-layer compound, with a simple tetragonal crystal structure and an optimal $T_c$ of nearly 100 K
\cite{Eisaki2004,Barisic2008,Li2011,BarisicPNAS2013,Barisic2013a,Mirzaei2013,Chan2014,Barisic2019,Chan2016a,ChanNComm2016,Popcevic2018,Pelc2019a,Pelc2019}.  
The model nature of Hg1201 is exemplified, e.g., by the 
\textcolor{black}{observations that the normal-state magnetoresistance in the PG state exhibits Kohler scaling \cite{Chan2014} and that} 
low-temperature transport measurements reveal Shubnikov-de-Haas oscillations due to Fermi-surface reconstruction associated with the CDW order \cite{Barisic2013a,ChanNComm2016}. 
The zero-field CDW \textcolor{black}{order measured via RXS exhibits short in-plane correlation lengths, and there is no evidence for a phase  transition} \cite{Tabis2014,Tabis2017}. Gaining further understanding of these emergent correlations, \textcolor{black}{and of their connection with the PG phenomenon and the pairing glue,} is the primary goal of the present work. The comparatively high onset temperatures, reduced dimensionality, and short length scales suggest that significant dynamic CDW correlations might be present, in analogy with well-known systems such as NbSe$_2$ in the presence of point disorder \cite{Berthier1978,Arguello2014,Wu2015,Chatterjee2015}. 
\textcolor{black}{On the other hand, the CDW phenomenology in the cuprates is quite different from conventional systems, as it occurs in the part of the phase diagram that is dominated by the highly unusual PG phenomenon.} 

RIXS is a unique probe to study such effects. 
Indeed, at 250 K, well above the temperature ($T_\mathrm{CDW} \approx 200$ K) below which  \textcolor{black}{evidence for quasistatic} short-range CDW order was previously observed in Hg1201 via RXS \cite{Tabis2014, Tabis2017} and time-resolved optical reflectivity \cite{Hinton2016}, we only discern $dynamic$ \textcolor{black}{charge} correlations, with a characteristic energy of about 40 meV. 
\textcolor{black}{At 70 K ($\approx T_c$), on the other hand, we observe quasistatic, short-range CDW correlations, and the dynamic response now includes an additional feature in the 100-250 meV range that is broadly peaked at about 165 meV. The response centered at about 40 meV does not appear to be affected on cooling from 250 K to $T_c$ and may be of mixed charge/phonon character.
These two energy scales appear in other observables, and have been associated with both the PG \cite{Honma2008} and the superconducting pairing glue \cite{Dordevic2005,Hwang2007,vanHeumen2009,Muschler2010,Ahmadi2011,dalConte2012}. 
Moreover, the larger of the two scales} has been linked to the strong electronic correlations that cause the charge-transfer gap of the undoped parent insulators \cite{Pelc2019a,Pelc2019}. 
\textcolor{black}{Our experiment therefore potentially detects not only CDW correlations, but charge correlations in a broader sense.
We demonstrate that energy integration of the new RIXS data yields the prior RXS result for Hg1201, including the high-temperature feature seen at $\boldsymbol{q}_\mathrm{CDW}$ in Hg1201 and other cuprates \cite{Tabis2014,Tabis2017}.
This high-temperature feature, which is typically treated as background, is therefore shown to be associated with dynamic charge correlations. 
From a comparison with RXS data for other single-layer cuprates at the doping level for which the CDW order is most robust, we find that the strength of the high-temperature dynamic signal is consistently 35-50 \% of the signal at $\sim T_\mathrm{c}$. 
Given the ubiquitous presence of the high-temperature feature in prior RXS data, even at doping levels where CDW correlations are weak, this points to the universal existence of significant dynamic charge (or mixed charge/phonon) correlations throughout the temperature-doping phase diagram.
We discuss the possibility that these charge correlations are rooted in the doped-charge-transfer-insulator nature of the cuprates. 
}
\textcolor{black}{The present data naturally yield the paramagnon dispersion along [1,0], which we find to be consistent with magnetic neutron scattering results \cite{Chan2016a,Tang2019}. The magnetic excitations are nearly unaffected by the unusual charge correlations, consistent with their somewhat different energy scales. However, comparison with the characteristic energy scales of the glue function extracted from optical and Raman spectroscopy points to the distinct possibility that both charge and magnetic excitations contribute to the superconducting pairing glue in the cuprates.}

\begin{figure}[t]
\hspace*{-2.5mm}\includegraphics[width=8.5cm]{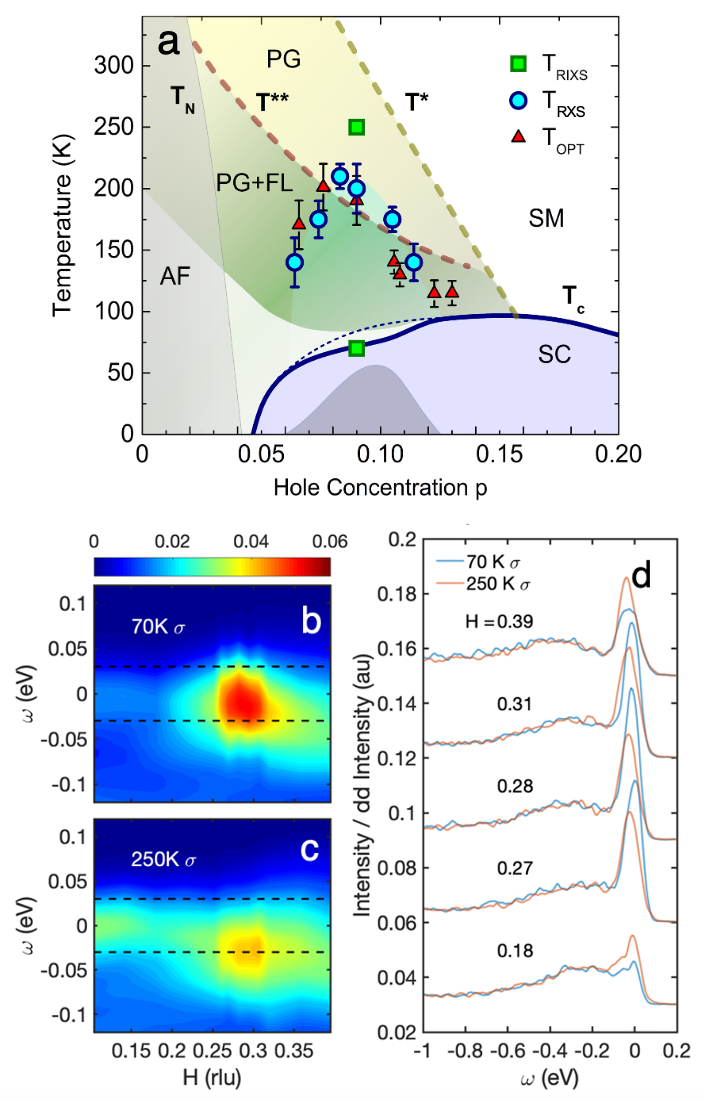}
\vspace{-3mm}
\caption{(a): Hg1201 phase diagram ($p>0.04$), extrapolated to $p=0$ based on results for other cuprates \cite{Keimer2015}. Solid blue line: Doping dependence of the superconducting (SC) transition temperature $T_c(p)$ \cite{Yamamoto2000}. Dark gray region: Deviation (not to scale) of $T_c (p)$ from estimated parabolic dependence (dashed blue line). $T^*$: Pseudogap (PG) temperature, estimated from deviation from strange-metal (SM) $T$-linear resistivity \cite{BarisicPNAS2013,Barisic2019}. $T^{**}$: Temperature below which Fermi-liquid (FL) behavior is clearly observed in the PG state \cite{BarisicPNAS2013,Mirzaei2013,Chan2014,Barisic2019}. $T_{RXD}$: Onset of short-range CDW correlations estimated from EI-RXS \cite{Tabis2014,Tabis2017}. $T_{OPT}$: Characteristic temperature observed in time resolved optical reflectivity \cite{Hinton2016}. $T_{RIXS}$: Temperatures at which the Hg1201 sample ($T_c = 70$ K, $p \sim 0.086$) was measured in the present work. (b) and (c): Momentum-energy contours of RIXS spectra at 70 K and 250 K, \textcolor{black}{obtained with $\sigma$ polarization}. Dashed lines: full-width-at-half-maximum (FWHM) energy resolution of 60 meV. (d) Spectra at select $\boldsymbol{H}$ values, \textcolor{black}{obtained with $\sigma$ polarization}; data are vertically shifted for clarity.}
\vspace{-4mm}
\label{Fig. 1}
\end{figure}
\section{Experimental Methods}
The measurements were performed with the ERIXS spectrometer at beam line ID32 of the European Synchrotron Radiation Facility (ESRF), Grenoble, France \cite{Brookes2018}. The incident X-ray energy was tuned to the maximum of the Cu $L_3$ absorption peak around 932 eV, and the X-ray polarization was set either parallel ($\pi$) or perpendicular ($\sigma$) to the scattering plane \cite{Supplemental}. The scattered photons were analyzed without considering the final-state polarization. The energy resolution was approximately 60 meV, as determined from the full width at half maximum (FWHM) of the non-resonant spectrum of a standard polycrystalline silver sample. 
In order to prepare a clean, high-quality surface, the Hg1201 single crystal was cleaved \textit{ex situ} to reveal a face parallel to the CuO$_2$ planes. 

\textcolor{black}{The cleaved surface was uniform and large compared to the beam size, thus minimizing the possibility of the beam drifting off the sample. In order to avoid effects due to beamline energy drifts, the resonant energy was periodically checked via X-ray absorption spectroscopy at the Cu $L$ edge. At each momentum transfer, the energy zero was determined by measuring the non-resonant spectrum of a polycrystalline silver standard located near the sample. The data were collected in short time intervals of 45 min, and each scan was corrected for the energy shift, if any was observed.
In order to avoid errors in the determination of the energy zero, the thickness of the silver was approximately the same as the thickness of the sample, and both were placed on the same axis of rotation. As a consistency check, we monitored the energy of $dd$-excitations and found no significant change with temperature in the effective $dd$ excitation energy, and hence in the energy zero \cite{Supplemental}.
Importantly, the energy resolution (60 meV, FWHM) and observed energy scales ($\sim 40$ meV and $\sim 165$ meV) are large compared to typical beamline energy drifts, thus minimizing the chance of an adverse affect on our conclusions.}

Momentum scans were performed by rotating the sample about the axis perpendicular to the scattering plane, and the detector angle was set to $2\theta = 150^{\circ}$. The scattering wave vector is $\boldsymbol{Q} = H\boldsymbol{a^*} + K\boldsymbol{b^*} + L\boldsymbol{c^*} = (H, K, L)$ in reciprocal lattice units, where $a^* = b^* = 1.62$ \AA$^{-1}$ and $c^* = 0.66$ \AA$^{-1}$. $K$ was chosen to be zero, and the scans were taken along $[H,0,L]$, with $L$ coupled to $H$. 
Due to the short-range nature of the observed two-dimensional (2D) correlations, the $L$ dependence of the cross section is expected to be negligible, and we quote the 2D reduced wave vector  $\boldsymbol{q} = (H, K)$.
The intensity of the RIXS spectra was normalized to the integrated intensity of $dd$ excitations, following prior work \cite{Braicovich2010}. 

\section{Results}
Figure 1(b,c) shows RIXS intensity contour plots at $T = 70$ K and 250 K, respectively, as a function of $H$ and energy transfer $\omega$ (negative $\omega$ corresponds to energy loss), obtained with $\sigma$-polarized incident X-rays directly from individual energy scans such as those shown in Fig. 1(d) \cite{Supplemental}. 
The maps clearly display signal around $\boldsymbol{q}_{CDW} \approx (0.28, 0)$, consistent with prior RXS work \cite{Tabis2014, Tabis2017}. 
Whereas the dominant signal at 70 K is quasielastic, at 250 K ($> T_\mathrm{CDW}$) the response is dynamic and centered in the optic phonon range.
Due to the larger phonon contribution, the 250 K spectra exhibit higher intensity away from $\boldsymbol{q}_{CDW}$  than the 70 K data. 
\textcolor{black}{Although weaker, as expected, charge} correlations are also observed in $\pi$ polarization \cite{Supplemental}.

\begin{figure}[t]
\hspace*{0mm}\includegraphics[width=7cm]{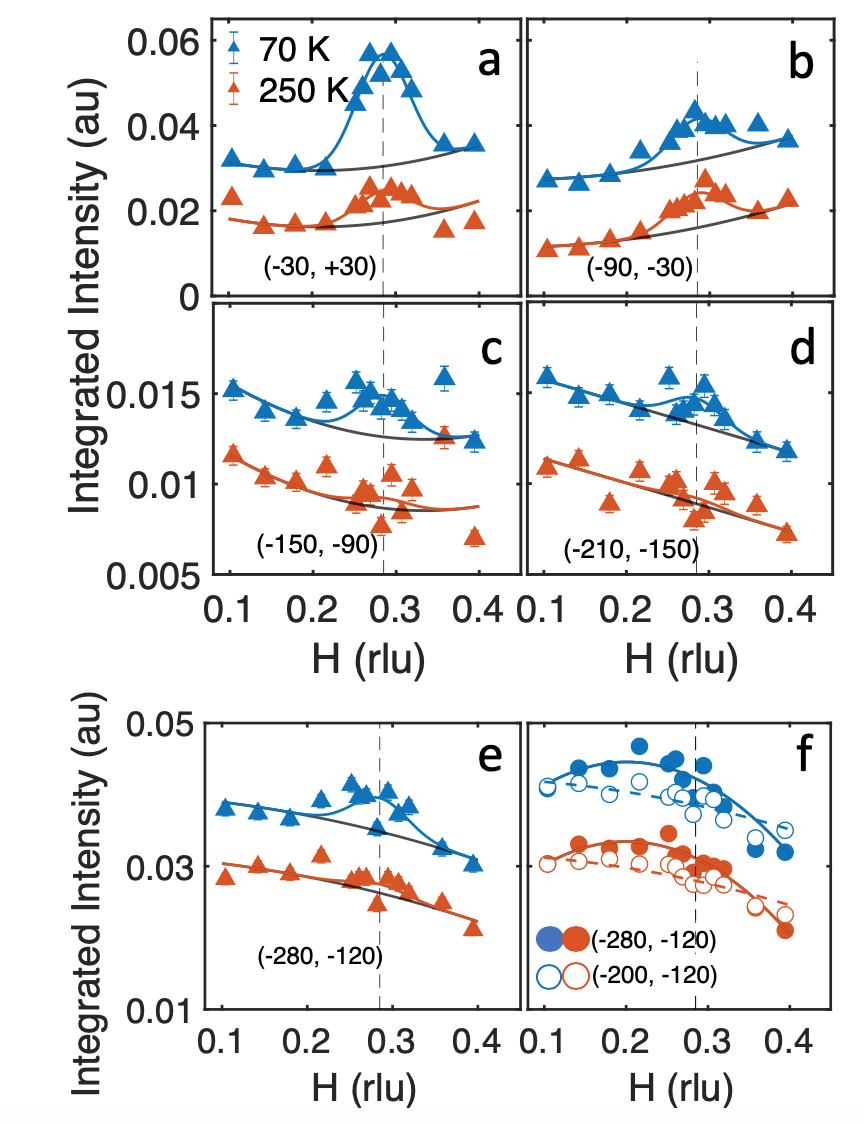}
\vspace{-3mm}
\caption{(a-d) $\sigma$-polarized RIXS intensity spectra at 70 K and 250 K, integrated over the \textcolor{black}{60 meV} FWHM instrument resolution, centered at zero, 60, 120 and 180 meV energy loss. (e) $\sigma$- and (f) $\pi$-polarized RIXS spectra at 70 K and 250 K, integrated over larger energy windows, as indicated. In all cases, the 70 K curves are vertically shifted for clarity. The black lines in (a-e) are polynomial momentum dependences, and the same in each case at low and high temperature. Blue and red lines in (a-e): fits to Gaussian \textcolor{black}{profiles plus a smoothly varying contribution} at 70 K and 250 K, respectively. Blue and red lines in (f): guides to the eye. Vertical dashed grey lines indicate $\boldsymbol{q}_{CDW}$. 
\textcolor{black}{The errors are one standard deviation (square root of total photon count).}}
\vspace{0mm}
\label{Fig. 2}
\end{figure}

A more detailed data analysis reveals additional information. Figure 2(a-d) shows the momentum dependence of RIXS signal (obtained from energy scans such as those in Fig. 1(d), with $\sigma$ polarization) integrated over the FWHM energy resolution (60 meV) with different energy ranges: quasielastic (-30, +30) meV; inelastic (-90, -30) meV, (-150, -90) meV, and (-210, -150) meV.  
In order to arrive at a systematic estimate of the \textcolor{black}{$q$-integrated} signal strength, we \textcolor{black}{perform a heuristic fit of} the data to a Gaussian peak, with fixed center ($\boldsymbol{q}_{CDW}$) and width ($\sim 0.075$ rlu), plus a concave, polynomial contribution. \textcolor{black}{For Hg1201, the optic phonon range extends to about 75 meV \cite{dAstuto2003,UchiyamaPRL2004,Reznik2019}, and for energy transfers in this range the} ``background" contribution invariably includes phonon scattering; for all energy-integration ranges, \textcolor{black}{this smoothly-varying contribution} is indistinguishable at 70 K and 250 K. 

\textcolor{black}{The ``elastic line" in inelastic scattering mainly originates from scattering off defects and disorder. Due to the high quality of our cleaved sample, the true elastic line is almost impossible to resolve in the raw spectra (Fig. 1d and Fig. S2 in \cite{Supplemental}), as the quasi-elastic signal is always contaminated by other low-energy excitations (mostly phonons). The intensity increase around zero energy and $\boldsymbol{q}_{CDW}$ due to quasistatic CDW correlations is clearly distinct from the elastic response. Our analysis indicates that the tail of the energy resolution vanishes faster than a Lorentzian function. The increase of the elastic line at small $H$ at 250 K (Fig. 2a) does not increase the inelastic signal (Fig. 2b), which further supports our assessment that the high-energy signal is genuine.} 

Figure 3(a,b) shows the energy dependence of the Gaussian amplitude obtained in this manner for $\sigma$- and $\pi$-polarization, respectively. \textcolor{black}{As expected for charge scattering, the signal is weaker for $\pi$-polarization \cite{Supplemental}.} The 250 K results show broad peaks centered at $\sim 40$ meV and no evidence for elastic scattering, as the energy dependence of the amplitude is fully captured by the sum of Stokes and anti-Stokes scattering. 
\textcolor{black}{The peaks are considerably broader than the 60 meV energy resolution. Assuming a heuristic intrinsic Gaussian profile, we estimate an intrinsic width of at least 50 meV (FWHM) after resolution deconvolution. The RIXS signal may be interpreted either as a single charge mode, with a width that is comparable to its characteristic energy, or as} the result of a distribution of charge modes. \textcolor{black}{It is equally well described by a log-normal distribution} \cite{Supplemental}.
\textcolor{black}{We note that our analysis removes phonon contributions that vary smoothly with $H$.} Whereas the broad peak in Fig. 1(c) contains \textcolor{black}{low-energy excitations, especially phonons, the extracted intensity at $\sim 40$ meV shown in Fig. 3(a,b) signifies an enhanced charge response at $\boldsymbol{q}_{CDW}$.
As discussed in detail below, this enhanced response may in part be due to anomalous phonon scattering. 
}

At 70 K, the dominant \textcolor{black}{charge} response is quasistatic. The width of this peak is larger than the energy resolution, consistent with a dynamic contribution at $\sim 40$ meV \textcolor{black}{that is unchanged from the response} at 250 K. This distinct possibility is highlighted for both polarizations in the (Bose-factor-corrected) intensity difference plots in Fig. 3(c,d), which reveal resolution-limited elastic peaks centered,  \textcolor{black}{within error,} at $\omega = 0$.

Interestingly, at 70 K we also observe dynamic charge fluctuations \textcolor{black}{well above the optic phonon range}. This is {\it directly} seen from Fig. 2, especially Fig. 2(e), where the large binning range (-280,-120) meV was chosen in order to optimize signal-to-background. These data, obtained with $\sigma$-polarization, are contrasted in Fig. 2(f) with the equivalent result with $\pi$-polarization, which is more sensitive to magnetic scattering \cite{Supplemental}. The convex momentum dependence can be attributed to paramagnons, which become prominent above $\sim 200$ meV at  \textcolor{black}{$\boldsymbol{q}_{CDW}$} (see below). This is seen from the comparison in Fig. 2(f) with the result obtained with narrower (-200,-120) meV integration, which yields an approximately linear background consistent with the $\sigma$-polarization result in Fig. 2(e). From Fig. 3(a), the high-energy charge signal is seen to be peaked at about  \textcolor{black}{165 meV}; it is not discerned in $\pi$-polarization (Fig. 3(b,d)) due to the higher background level (proximity to paramagnon excitations; Fig. 2(f)) and lower expected charge scattering cross section (by a factor of two \cite{Supplemental}).
\textcolor{black}{We note that the 250 K data in Figs. 2  and 3(a) are consistent with non-zero high-energy charge signal, although with a smaller amplitude and potentially larger momentum width than at 70 K.
}

\begin{figure}[t]
\begin{center}
\hspace*{-4mm}\includegraphics[width=9.5cm]{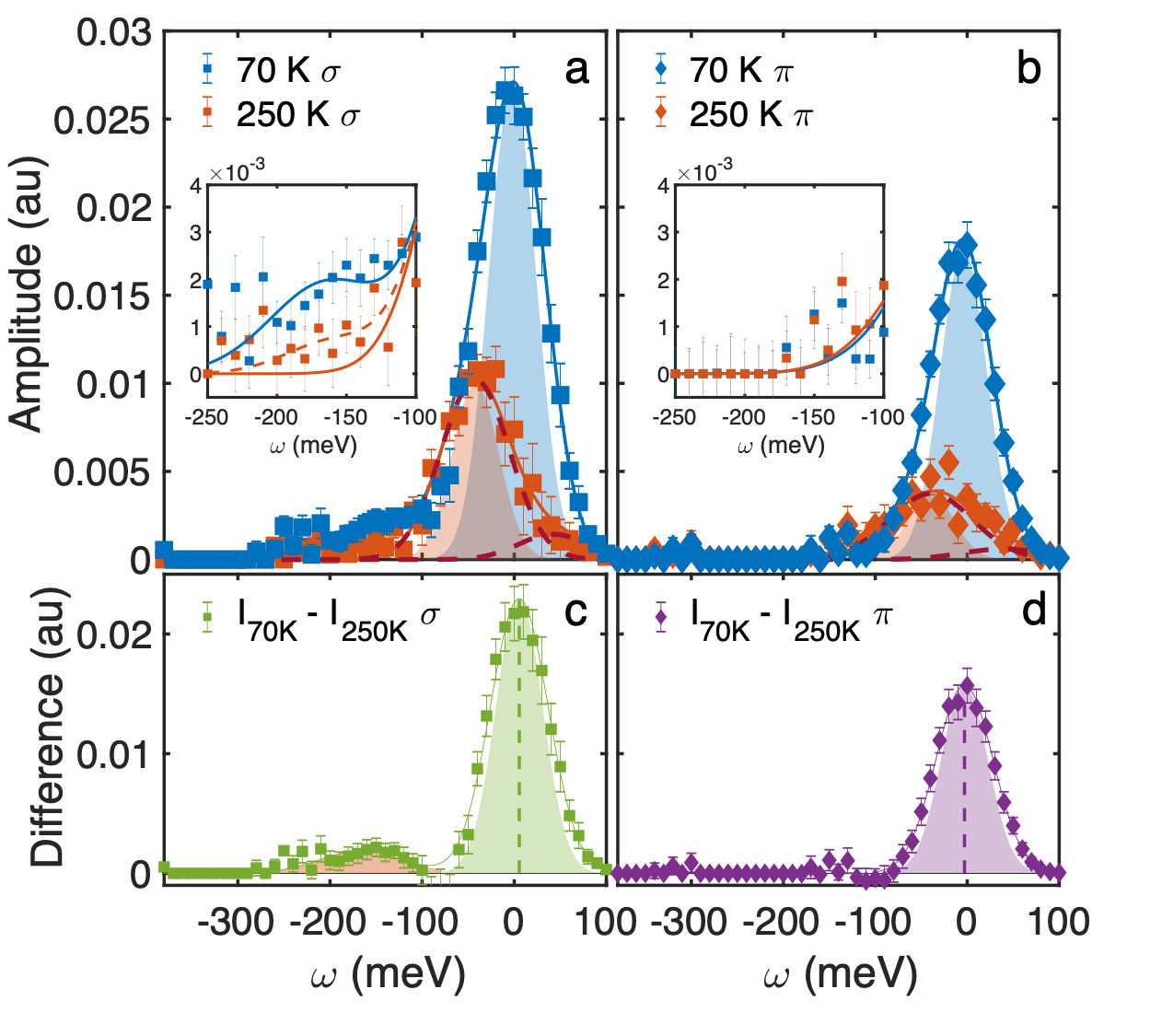}
\end{center}
\vspace{-9mm}
\caption{Energy dependence for (a) $\sigma$- and (b) $\pi$-polarization of fitted Gaussian amplitude from energy-integrated data such as those shown in Fig. 2(a-d). \textcolor{black}{The error bars indicate fit errors.} Insets: zoom of the range (-250, -100) meV; \textcolor{black}{the dashed red line in the inset to (a) indicates the possibility of non-zero charge scattering in this energy range at 250 K.}
Blue and red shaded areas indicate the instrument resolution of 60 meV (FWHM). Gaussian fits to Stokes and anti-Stokes scattering at 250 K (dashed red lines; sum: solid red lines) yield \textcolor{black}{41(4)} meV and \textcolor{black}{64(12)} meV for the peak position and intrinsic (de-convoluted) FWHM peak width. Gaussian fits to three peaks at 70 K (blue lines) capture (1) quasistatic, (2) low-energy ($\sim 40$ meV), and (3) high-energy (peak at \textcolor{black}{163(12)} meV, intrinsic width of \textcolor{black}{77(16)} meV (FWHM)) contributions to the CDW response; the latter is not discerned in $\pi$-polarization, for which the charge response is expected and seen to be weaker \cite{Supplemental}.  
(c) and (d): difference in amplitude between 70 K and 250 K from (a) and (b), respectively, after correcting the 250 K data for the Bose factor. Green and purple shaded areas indicate the instrument resolution; vertical dashed lines indicate peak centers obtained from fits to Gaussian profiles, which are consistent within error with zero energy transfer. The orange shaded area in (c) indicates the net response centered at 160(6) meV.}
\vspace{0mm}
\label{Fig. 3}
\end{figure}

Figures 1(d) and 4(a) reveal an additional broad, dispersive peak in the 0.15 to 0.4 eV range. 
This feature, which has been observed in a number of cuprates, signifies paramagnon scattering that evolves from well-defined antiferromagnetic excitations in the undoped parent compounds \cite{Haverkort2010, Ament2009}. These excitations are more prominent in $\pi$- than in $\sigma-$polarization in the present scattering geometry, as expected for magnetic scattering \cite{Supplemental,LeTacon2011, Murakami2017}. 
We extract the paramagnon energy by fitting the raw RIXS spectra to a resolution-limited elastic peak, \textcolor{black}{an effective} phonon peak, and a damped paramagnon excitation \cite{Supplemental}.  
Figure 4(b) summarizes our result for the paramagnon dispersion along [1,0] at 70 K. 
We compare the RIXS data with magnetic neutron scattering data near the antiferromagnetic wave vector for two Hg1201 samples (one with essentially the same doping level and $T_c = 71$ K \cite{Chan2016a}, and the other with $p \approx 0.064$ and $T_c = 55$ K \cite{Tang2019}) and find that these  \textcolor{black}{results} are highly consistent and complementary. From a heuristic fit of the combined neutron and X-ray data above $H=0.1$ rlu to simple linear spin-wave theory, we obtain an effective nearest-neighbor exchange of 123(3) meV. \textcolor{black}{The relatively large uncertainty and limited data near the zone boundary prevent a more detailed analysis \cite{Coldea2001, Headings2010, Ivashko2017}.}
Overall, the RIXS data for the paramagnon dispersion in Hg1201 are consistent with prior measurements for other hole-doped cuprates \cite{LeTacon2011, Dean2013a, Ivashko2017, Supplemental}.

\begin{figure}[t]
\hspace*{0mm}\includegraphics[width=8.5cm]{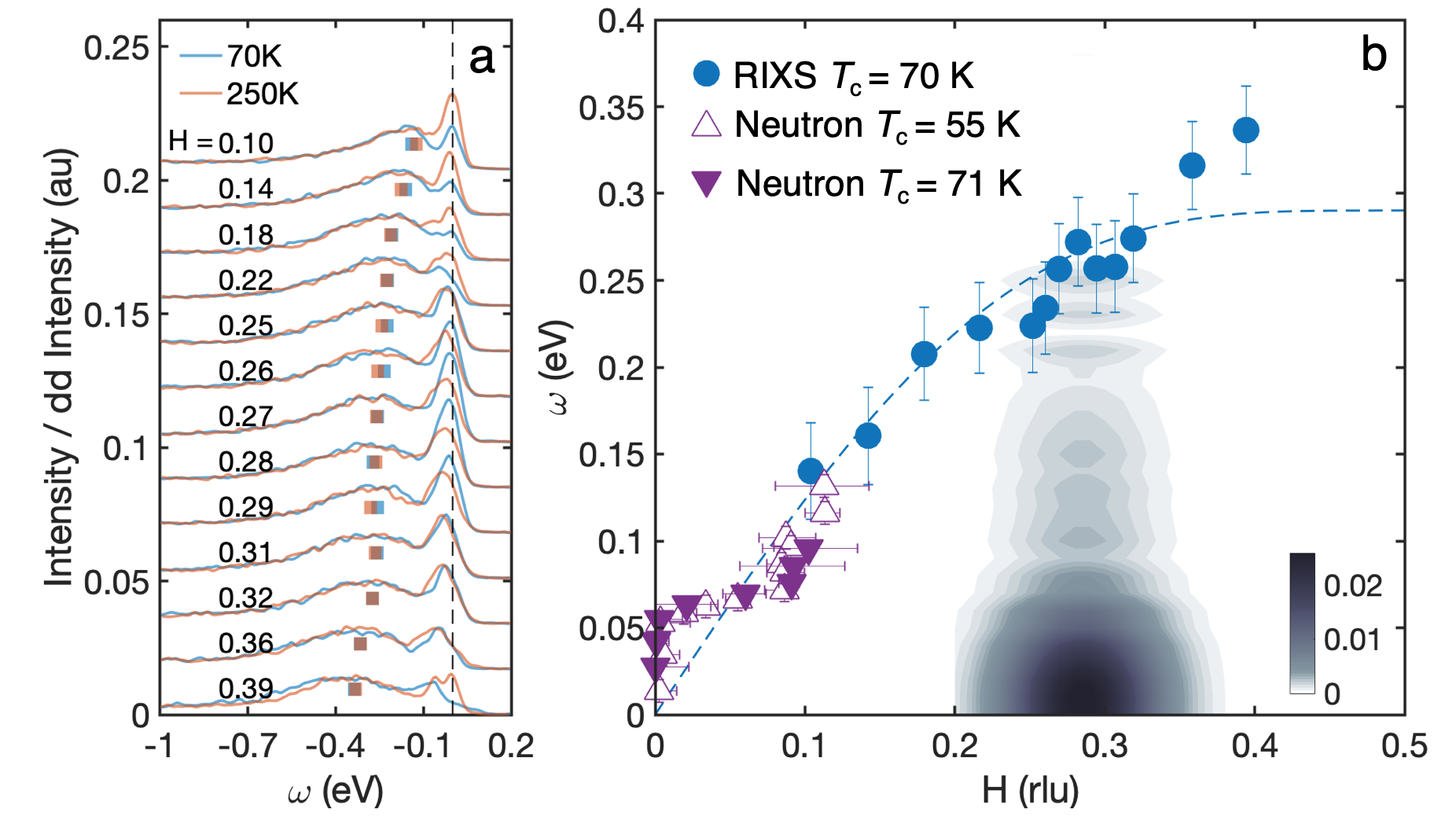}
\vspace{-3mm}
\caption{(a). RIXS spectra for Hg1201 at numerous wave vectors along [1,0], obtained with $\pi$-polarization. Data are vertically shifted for clarity. Blue and red bars indicate the paramagnon energy obtained from fits to a damped harmonic oscillator (see text and \cite{Supplemental}). (b) Dispersion of magnetic excitations in Hg1201 measured by RIXS (this work; $p = 0.086, T_c = 70$ K) and neutron scattering for two Hg1201 samples: one with nearly the same doping level and $T_c = 71$ K \cite{Chan2016a}, and the other with a slightly lower doping level of $p \approx 0.064$ and $T_c = 55$ K  \cite{Tang2019}. Blue dashed line: heuristic fit to nearest-neighbor spin-wave theory (see text). Error bars for RIXS data are set to 30 meV, i.e., half of the FWHM energy resolution. The contour indicates the relative charge intensity at 70 K (Fig. 3(a)).}
\vspace{0mm}
\label{Fig. 4}
\end{figure}
\section{Discussion}

\textcolor{black}{
We observe charge signal in three different energy ranges: (1) a quasistatic response at 70 K; (2) low-energy ($\sim 40$ meV) \textcolor{black}{signal} at 250 K and 70 K; (3) high-energy fluctuations in the 100 - 250 meV range at 70 K. 
Below, we first discuss how our observations differ from conventional CDW systems. We then consider the possibility that the 40 meV feature signifies a phonon-related effect. 
Finally, we discuss our findings within the broader context of the cuprate phase diagram and, in particular, the PG phenomenon and the superconducting pairing glue.
}

\subsection{\label{sec:level2} \textcolor{black}{Cuprates vs. conventional CDW systems}}

\textcolor{black}{Analogous to conventional superconductors, conventional CDW systems are described by BCS theory \cite{Kohn1970}. The CDW state is a condensate of electrons and holes, characterized by a complex order parameter, whose net momentum corresponds to the wavelength of the charge order. In clean, weak-coupling systems, macroscopic phase coherence and amplitude formation occur simultaneously at the mean-field transition temperature. In the presence of disorder or at strong coupling, on the other hand, the transition temperature can be significantly depressed, and such systems may exhibit extended PG-like behavior, with a gap that persists to high temperature in the absence of long-range order. Such an extended regime has been demonstrated for the canonical CDW compound $2H-$NbSe$_2$ intercalated with Mn and Co, i.e., in the presence of point disorder \cite{Berthier1978,Arguello2014,Wu2015,Chatterjee2015}.}

\textcolor{black}{
The PG and charge-order behavior of the cuprates are more complex. These compounds transform from Fermi-liquid metal at high doping, with a large Fermi surface that encompasses $1 + p$ carriers, to a Mott insulator at zero doping ($p=0$) via an intermediate state with carrier density $p$ \cite{PadillaPRB2005,OnoPRB2007,BarisicPNAS2013}. This evolution involves the localization of exactly one hole per planar CuO$_2$ unit and the concomitant opening of a PG on the ``antinodal" parts of the underlying Fermi surface, which leaves ungapped nodal Fermi arcs with carrier density $p$ below $T^{**}$ (Fig. 1), the lowest of three characteristic PG temperatures (the other two are $T^*$ and $T_{hump}$, with $T^{**} < T^* < T_{hump}$) \cite{Honma2008,Pelc2019a}. This Mott-pseudogap phenomenon is distinct from the behavior of a canonical CDW system in the presence of strong coupling and/or disorder \cite{Chatterjee2015}. The onset of short-range \textcolor{black}{CDW correlations} is generally observed at or below $T^{**}$ (Fig. 1). Associated with the three characteristic PG temperatures are characteristic energy scales, which at the doping level of the present study are about 40-50, 60-80, and 180-220 meV \cite{Honma2008,Alldredge2013}. The lowest of these scales is the value of the PG near the tip of the arcs \cite{Alldredge2013}, whereas the largest scale has been associated with the charge-transfer gap of the insulator, renormalized by the itinerant carriers \cite{Pelc2019a,Pelc2019}. Since the PG formation is gradual and precedes the CDW order \cite{Gomes2007,Tallon2019}, and because the wavevector $\boldsymbol{q}_{CDW}$ appears to connect the tips of the arcs \cite{Tabis2014,ChanNComm2016,Tabis2017}, the 40-50 meV PG scale likely sets an upper bound for the CDW gap scale.  
From the BCS expression $2\Delta = 3.5 k_B T_{CDW}$, with $T_{CDW} \sim 200$ K (Fig. 1), one would expect a gap of $\Delta \sim $ 30 meV.   
Given the considerable point disorder exhibited by the cuprates \cite{Eisaki2004}, it is perhaps not surprising to observe CDW correlations over an extended temperature range below $T_{CDW}$, along with a dynamic signature above this nominal ordering temperature, which in this scenario represents a pinning temperature rather than a true phase transition.
}

\textcolor{black}{However, in this case one would expect a temperature-dependent dynamic scale and a transfer of spectral weight from the dynamic to the static response on cooling, yet this is not what we observe. The difference plots in Fig. 3(c,d) indicate that the Bose-factor corrected weight of the $\sim40$ meV} inelastic contribution is temperature-independent, since it cancels out within error for the two temperatures. Equivalently, there seems to be little spectral weight transfer from the $\sim 40$ meV dynamic to the quasistatic contribution on cooling, which would not be expected if the CDW were simply pinned below $T_{CDW}$. Therefore, the quasistatic and $\sim 40$ meV contributions likely have different physical origins.
\textcolor{black}{In principle, one possibility is an elastic contribution due to} the interference of Friedel oscillations around impurities, as suggested theoretically \cite{DallaTorre2016} and tested in a quasi-one-dimensional CDW system \cite{Yue2019}. The true underlying CDW signal would then be inelastic even at \textcolor{black}{$T_c$}, and the CDW correlations would remain dynamic at this temperature. 
\textcolor{black}{Yet even in this case one would expect a temperature-dependent dynamic CDW signal. A more likely possibility is that the CDW response is quasielastic at all temperatures, whereas the dynamic $\sim 40 $ meV signal is of a different origin. This conclusion is supported by recent work for NdBa$_2$Cu$_3$O$_{6+\delta}$ (NBCO) \cite{Arpaia2019}, which found the amplitude and correlation length of the $\omega \approx 0$ signal to exhibit temperature dependences below $T_{CDW}$ that point to a gradual build-up of \textcolor{black}{quasistatic} correlations and to an underlying phase transition at a temperature below $T_c$.  
As further discussed in the following subsections, it is therefore a distinct possibility that the temperature-independent (at least up to 250 K) $\sim 40$ meV feature at $\boldsymbol{q}_\mathrm{CDW}$ is 
either a direct signature of some other (non-CDW) bosonic charge modes, or the result of anomalous phonon scattering, and hence an indirect signature of such modes. 
}   

\textcolor{black}{In a conventional system, CDW order involves Fermi-surface nesting. 
The present data do not allow us to discern if the CDW in the cuprates} forms predominantly due to nesting or a real-space mechanism. We can, however, make some inferences at this point. The doping dependence of the CDW wave vector is consistent with a nesting scenario, and quantum oscillation measurements show that the Fermi surface is reconstructed at low temperatures and high magnetic fields, with a resultant electron pocket whose size is consistent with a simple reconstruction scenario \cite{Barisic2013a,ChanNComm2016}. However, the quasistatic short-range CDW correlations observed here \textcolor{black}{at $T_c$ clearly do not induce a reconstruction, since transport properties are virtually insensitive to the CDW formation at this temperature \cite{Doiron-Leyraud2013}.} Furthermore, the presence of a significant dynamic component with high onset temperature and the emergent high-energy scale, \textcolor{black}{potentially both signatures of the PG,} suggest a strong-coupling scenario, making an underlying real-space mechanism more likely. This is to be expected if the CDW is indeed an emergent phenomenon, since the correlations that cause the cuprate PG appear to be local in real space, \textcolor{black}{as further discussed below}. It seems possible that, effectively, both $k$- and $r$-space effects contribute, which could be related to the deeper question of the existence of a well-defined reciprocal space in a material that is inherently inhomogeneous at the nanoscale \cite{Krumhansl1992,Gomes2007,Alldredge2013,Pelc2019a} 

\subsection{\label{sec:level2}\textcolor{black}{Nature of the 40 meV excitation}}

\textcolor{black}{The 40 meV scale} is consistent with superconducting gap and PG scales for cuprates with a comparable optimal $T_c$ \cite{Honma2008,Alldredge2013}, and with the lower bound of $\sim 20$ meV for the gap between the \textcolor{black}{CDW-reconstructed pockets} deduced from quantum oscillation experiments for Hg1201 \cite{ChanNComm2016}.
\textcolor{black}{This scale lies in the optic phonon range and, in principle, the observed excitation might signify phonon scattering due to enhanced electron-phonon coupling, and thus be an $indirect$ signature of significant dynamic charge correlations. 
Although there is no evidence for anomalous phonon scattering at $\boldsymbol{q}_\mathrm{CDW}$ off resonance (see Fig. 5b for a prior RXS result for Hg1201 \cite{Tabis2014}), anomalous scattering involving valence electrons could be significantly enhanced at the Cu $L$ edge.
In hole-doped Bi$_2$Sr$_2$CaCu$_2$O$_{8+\delta}$ (Bi2212) \cite{Chaix2017}, charge correlations were identified up to about 60 meV. These correlations were suggested to be dispersive and associated with strong Fano interference at the intersection with the Cu-O bond-stretching phonon. At low temperature, a gradual phonon softening near $\boldsymbol{q}_\mathrm{CDW}$ was observed, with a minimum energy of 40-45 meV and an intrinsic width of 30-40 meV (FWHM).  
In Hg1201, at $\boldsymbol{q}_\mathrm{CDW}$, the Cu-O bond-stretching phonon lies in the 55-60 meV range, whereas the Cu-O bond-bending phonon lies in the 40-45 meV range \cite{dAstuto2003,UchiyamaPRL2004,Reznik2019}. 
It is therefore possible that a similar anomaly involving predominantly the bond-bending phonon is present in Hg1201.
}

\textcolor{black}{ 
Alternatively, the 40 meV feature could be a $direct$ signature of charge fluctuations, for a number of reasons. 
First, unlike the result for Bi2212 \cite{Chaix2017}, our data indicate no change in the $\sim 40$ meV charge response (amplitude, characteristic energy and width) between high and low temperatures (Fig. 3). Second, we estimate the $intrinsic$ width of the excitation to be at least 50 meV (FWHM) (Fig. 3). This value is considerably larger than what might be associated with phonon linewidth broadening, and about 50-100\% larger than the width of the peak ascribed to anomalous phonon scattering in Bi2212 \cite{Chaix2017}. 
Third, we note that conventional RXS measurements generally reveal a local intensity maximum at/near $\boldsymbol{q}_\mathrm{CDW}$ already at high temperatures \cite{Ghiringhelli2012, Tabis2014, Comin2014, daSilvaNeto2016, daSilvaNeto2014}. This feature is commonly regarded as a temperature-independent background and subtracted from the low-temperature data to extract the CDW signal, although it also has been associated with charge correlations \cite{daSilvaNeto2016,Chaix2017}, as clearly supported by our RIXS results in Figs. 2 and 3.
By integrating over the whole energy range (-10, 2) eV of our experiment, we demonstrate in Fig. 5(a,b) that our RIXS data are highly consistent with the previous RXS study of Hg1201 at the same doping level \cite{Tabis2014}. As seen from Figs. 2 and 3, and further discussed in \cite{Supplemental}, most of the signal at/near $\boldsymbol{q}_\mathrm{CDW}$ originates from within the (-0.1,0.1) eV energy range.  
In Fig. 5c and 5d, we reproduce the corresponding RXS results for Bi$_2$Sr$_{2-x}$La$_x$CuO$_{6+\delta}$ (Bi2201, $x=0.115$) \cite{Comin2014} and electron-doped Nd$_{2-x}$Ce$_x$CuO$_4$ (NCCO, $x=0.145$) \cite{daSilvaNeto2016}, respectively. Like Hg1201, both of these cuprates are single-CuO$_2$-layer compounds. The chosen doping levels are approximately those at which the CDW phenomenon is most robust. Remarkably, the ratios of the intensity amplitudes at/near $T_c$ and $T_{CDW}$ lie in the narrow 35-50\% range for all three single-layer cuprates.
Along with the insights gained here for Hg1201 (Figs. 2 and 3), this points to the existence of universal dynamic charge correlations, in addition to quasistatic correlations that emerge below $T_{CDW}$. 
Since phonon dispersions vary among the cuprates, and because any anomalous electron-phonon coupling can be expected to vary as well, this appears to rule out the possibility that the 40 meV feature we observe for Hg1201 is dominated by anomalous phonon scattering. Finally, we note that there exists indirect evidence for collective charge modes from neutron and X-ray scattering experiments of phonon anomalies throughout the superconducting doping range  
\cite{Reznik2012,ParkPRB2014}.
}

\textcolor{black}{ 
The most likely scenario is that there exist both types of contributions.
This is supported by a time- and frequency-resolved optical spectroscopy measurement of optimally-doped Bi2212 that revealed bosonic modes of predominantly electronic character in the optic phonon range, along with a significant strongly-coupled phonon contribution and a small lattice contribution \cite{dalConte2012}. 
The above observations point to universal dynamic charge correlations that extend to high energy and temperature, and that have little temperature dependence between room temperature and $T_c$.
Moreover, given that prior RXS measurements have revealed a local intensity maximum at/near $\boldsymbol{q}_\mathrm{CDW}$ even at high and low doping levels where there is no evidence for quasistatic, low-temperature CDW correlations
\cite{BlancoCanosaPRB2014,Supplemental}, we conclude that there exist universal dynamic charge correlations in a very large portion of the temperature-doping phase diagram. As discussed in the next subsection, this conclusion is further supported by indications from Raman spectroscopy of underlying dynamic charge modes that may, in fact, extend throughout the entire Brillouin zone.
These robust dynamic charge correlations therefore appear to be a precursor of the quasistatic CDW correlations that gradually develop below the doping-dependent temperature $T_{CDW}$ in a relatively narrow portion of the phase diagram, and that ultimately cause Fermi-surface reconstruction at very low temperatures, once superconductivity is suppressed with a $c$-axis magnetic field. 
}

\begin{figure}[t]
\hspace*{0mm}\includegraphics[width=8.5cm]{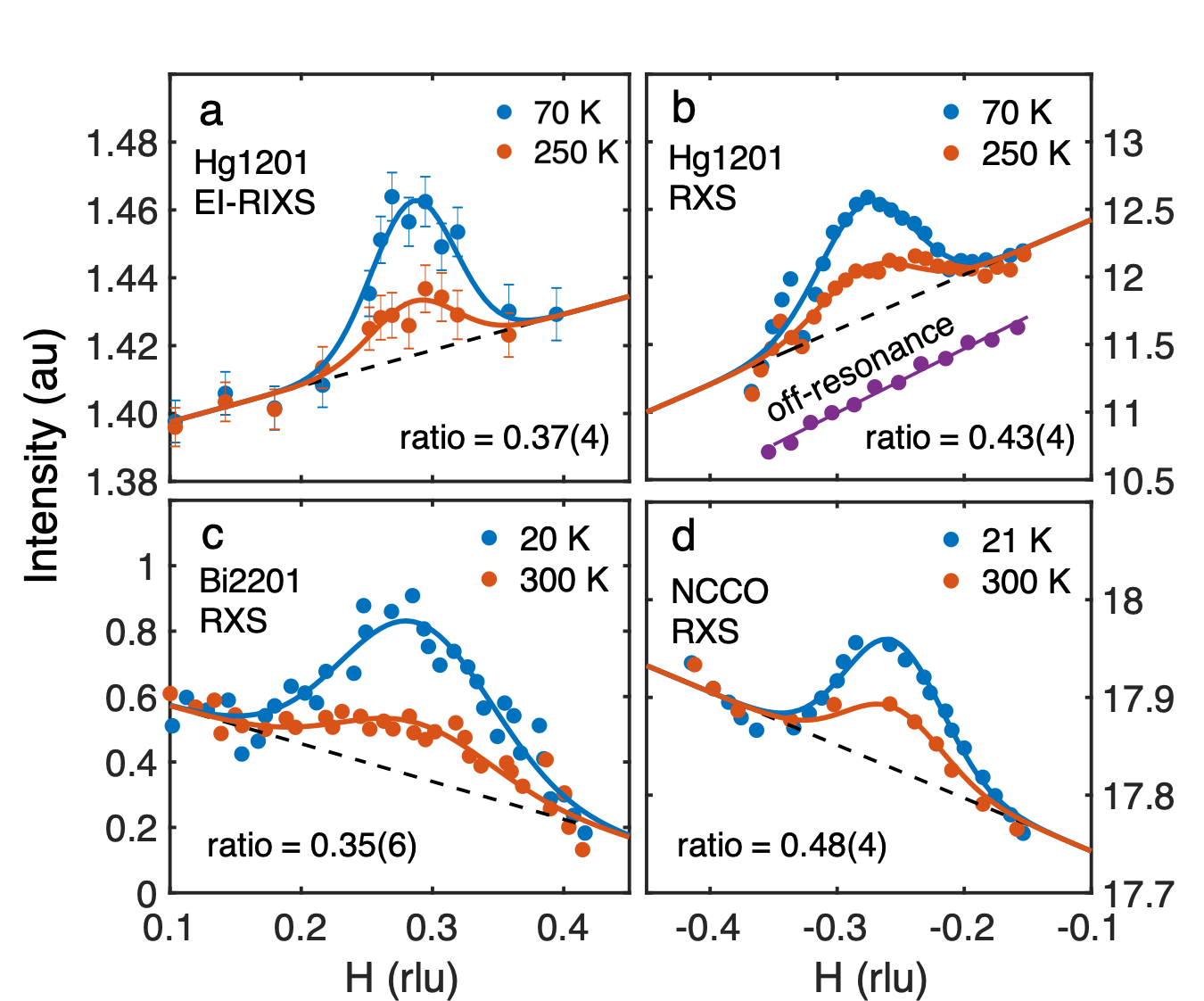}
\vspace{-3mm}
\caption{\textcolor{black}{(a) Energy-integrated RIXS (EI-RIXS) data for Hg1201, integrated over the entire experimental energy range (-10, 2) eV, both at $T_c$ and 250 K. (b) Prior RXS data for Hg1201 at the same doping level and temperatures as in (a), along with 70 K data taken off resonance with an incident photon energy of 929 eV, reproduced from \cite{Tabis2014}. (c) RXS measurement of Bi$_2$Sr$_{2-x}$La$_x$CuO$_{6+\delta}$ (Bi2201, $x=0.115$) at $\sim T_c$ and 300 K, reproduced from \cite{Comin2014}. (d) RXS results for superconducting electron-doped Nd$_{2-x}$Ce$_x$CuO$_4$ (NCCO, $x=0.145$) at $\sim T_c$ and 300 K, reproduced from \cite{daSilvaNeto2016}. In all cases, the high-temperature data are vertically shifted for comparison with the low-temperature result. Black dashed lines: assumed linear background. Solid lines: fits to Gaussian peak plus assumed linear background. Estimates of relative peak amplitudes are indicated in all panels.}}
\vspace{0mm}
\label{Fig. 5}
\end{figure}

\subsection{\label{sec:level2}\textcolor{black}{Comparison with other results for the cuprates}}

\textcolor{black}{RIXS measurements of} high-energy charge correlations above the optic phonon range have been reported so far only for electron-doped NCCO, where they persist up to approximately 0.4 eV \cite{daSilvaNeto2018}.
\textcolor{black}{Figures 2 and 3 demonstrate for Hg1201 at the investigated (hole) doping level that the full dynamic response peaked at $\boldsymbol{q}_\mathrm{CDW}$ is captured by integrating up to about 0.3 eV. 
We estimate that at least 20\% of the dynamic charge response at 70 K originates from above the optic phonon range \cite{Supplemental}.}

\textcolor{black}{A recent RIXS study of hole-doped NBCO \cite{Arpaia2019} found dynamic charge correlations to persist to high temperatures (above $T^*$). The experiment was performed with 40 meV (FWHM) energy resolution, and the data were analyzed under the assumption that phonon contributions at  {\bf q} = ($H$,0) can be removed by subtracting data obtained at the same $q$ along {\bf q} = ($H$,$H$), and by considering intensity differences for data obtained at different temperatures.
Dynamic energy scales of 15 meV (at 150 K and 250 K) and 7 meV (90 K) were deduced for an optimally-doped sample with $T_c = 90$ K ($p \approx 0.17$), whereas scales of 6 meV (150 K and 250 K) and 3 meV (90 K) were obtained for an underdoped sample with $T_c = 60$ K ($p \approx 0.17$). 
The high-temperature signal was found to exhibit short spatial correlations and to be centered at a somewhat smaller wave vector than the quasistatic CDW response. Our data for Hg1201 are overall consistent with these findings, but do not allow us to discern an additional $\sim 10$ meV dynamic signal due to the relatively large energy resolution of 60 meV (FWHM) of our experiment. In turn, the NBCO data are consistent with our observation of significant dynamic charge correlations at higher energies \cite{Supplemental}.
}  
 
The large $\sim 165$ meV energy scale identified here for hole-doped Hg1201 is consistent with the high-energy PG (``hump") scale seen in other observables \cite{ Honma2008,Pelc2019a}, \textcolor{black}{including the mid-infrared peak seen in optical spectroscopy experiments} \cite{Uchida1991,Lee2005}. 
A recent phenomenological model rooted in the dual empirical observations of a universal transport scattering rate \cite{Li2016,Barisic2019} and inherent inhomogeneity \cite{Pelc2018,Popcevic2018,Pelc2019c}, associates this scale with the charge-transfer gap at zero doping and with the delocalization of one hole per planar CuO$_2$ unit above $T^{**}$ (Fig. 1) \cite{Pelc2019a,Pelc2019}. 
For Hg1201, at $p \approx 0.09$, the mean (de)localization gap \textcolor{black}{and (Gaussian) gap distribution width that best capture the transport data are $\sim 180$ meV and $\sim 60$ meV (FWHM) \cite{Pelc2019a,Pelc2019}, respectively,} consistent with the characteristic CDW scale \textcolor{black}{of 163(12) meV} and width \textcolor{black}{of 77(16) meV established in the present work} (Fig. 3(a,c)).

\textcolor{black}{The gradual} delocalization of one hole per CuO$_2$ unit is a large effect that clearly manifests itself in, e.g., the evolution of the Fermi-surface with doping and the strong temperature and doping dependence of the Hall number \cite{OnoPRB2007,Pelc2019a,Pelc2019}. 
In contrast, the CDW order involves a \textcolor{black}{relatively} small fraction of a hole per unit cell. 
While X-ray scattering experiments cannot provide the absolute CDW amplitude, a value of about 
0.03 hole per CuO$_2$ unit was estimated for La$_{2-x}$Ba$_x$CuO$_4$ \cite{Abbamonte2005}. A separate, consistent estimate can be obtained from NMR: 
from the universal relation between oxygen hole content and NQR frequency \cite{Jurkutat2014} and
the $^{17}$O NQR line broadening in the CDW phase \cite{Wu2015}, we estimate 0.028 hole in the Cu $3d$ orbital for Hg1201 \cite{Reichardt2016}.
\textcolor{black}{Although $\sim 0.03$ hole per CuO$_2$ unit is a relatively large fraction of the nominal doping level, it is small compared to 1.}  
It therefore seems likely that the \textcolor{black}{CDW phenomenon in the cuprates} is a secondary, emergent phenomenon related to the strong correlations that underlie the hole localization. 
This is supported by STM evidence for a qualitative change in the CDW form factor of Bi2212 at a characteristic scale comparable to the PG scale \cite{Hamidian2016}.
Furthermore, it is known that the cuprates are intrinsically inhomogeneous, with local gaps \textcolor{black}{that vary at the nanoscale} and persist well above the PG temperature \cite{Gomes2007,Tallon2019}. 
Puddles of localized charge that sustain significant dynamic correlations thus may already exist outside of the nominal PG region (above $T^{**}(p)$). 
This could account for the present observation for Hg1201 at 250 K and for the similar result for NBCO \cite{Arpaia2019}. 
\textcolor{black}{The relevant wavevector would span the antinodal regions and hence be close to, but smaller than $q_\mathrm{CDW}$.} 

\textcolor{black}{We have argued that the high-temperature peak at/near ${\bf q}_{CDW}$ seen universally in the cuprates is dominated by dynamic charge fluctuations, which for Hg1201 at the studied doping level have a characteristic energy of $\sim 40$ meV. Although not apparent from the data summarized in Fig. 5, there exists evidence from RXS that dynamic charge correlations centered at/near ${\bf q}_{CDW}$ extend throughout a much larger portion of the Brillouin zone, with real-space correlations no larger than 1-2 lattice constants and little doping and temperature dependence (unlike the quasistatic CDW signal). This is perhaps most clearly seen from RXS measurements of YBa$_2$Cu$_3$O$_{6+\delta}$ \cite{BlancoCanosaPRB2014}, for which we estimate the approximately doping-independent 2D-integrated strength of the high-temperature signal to be at least an order of magnitude larger than the strongly temperature-dependent quasistatic CDW signal (at the doping level where CDW correlations are strongest) \cite{Supplemental}, consistent with a separate estimate based on the recent RIXS data for NBCO \cite{Arpaia2019}. 
In other words, whereas it appears from data such as those in Fig. 5 that the energy-integrated high-temperature signal is about 35-50\% of the low-temperature response, it may in fact be dramatically larger. 
This signal might be a direct signature of the fluctuations associated with the localized hole \cite{Pelc2019a,Pelc2019}, and the characteristic wavevector may correspond to the distance between the antinodal regions of the Fermi surface on which the PG develops, and hence be somewhat smaller than $q_{CDW}$. The 2D-integrated strength of the signal is still considerably smaller than what would be expected if it were fully associated with the localization of one hole per CuO$_2$ unit: naively, given that the CDW amplitude corresponds to about 0.03 hole per CuO$_2$ unit, one might expect this signal to be $(1/0.03)^2 \sim 10^3$ times larger than the quasistatic CDW response. It is possible that there exists a considerable incoherent (local in real space) contribution to the charge response up to 0.3 eV and even higher energies that is not captured by RXS and RIXS experiments. Similarly, the $\sim 165$ meV scale might not be seen at 250 K in our experiment because the high-energy charge fluctuations are incoherent at high temperatures. 
These observations are consistent with the fact that, whereas in optical spectroscopy experiments the spectral weight of the mid-infrared feature is relatively small \cite{Uchida1991,Lee2005} and the coherent contribution at temperatures below $T$** corresponds to $p$ \cite{Mirzaei2013}, integration up to about 2.5 eV (on the order of the charge-transfer gap) approaches $1 + p$, where 1 is attributed to the localized charge \cite{Mirzaei2013,Barisic2019}.
We note that recent momentum-resolved electron-scattering measurements of the charge fluctuations in Bi2212 revealed a featureless continuum up to $\sim 1$ eV \cite{Husain2019} for a wide range of hole concentrations, from underdoped to overdoped, consistent with the existence of significant local excitations.
}

\textcolor{black}{The above considerations are supported by neutron scattering, Raman scattering, infrared spectroscopy, and tunneling results. 
First, we note that evidence for an underlying collective charge mode has been deduced from anomalies in the Cu-O bond-stretching vibration in the 65-85 meV range single-layer La$_{2-x}$Sr$_x$CuO$_4$ (LSCO) in the superconducting doping range \cite{ParkPRB2014}. 
An anomalous line-width broadening was observed in the entire momentum range from the 2D zone center to the 2D zone boundary at (0.5,0) rlu. Given the large intrinsic width of the $\sim 40$ meV feature in Hg1201, these findings for LSCO are consistent with anomalous electron-phonon coupling and a potentially universal spectrum of bosonic charge modes.} Second, the broad $\sim 165 $ meV feature seen in our experiment appears to be related to features seen in Raman spectra with $B_{1g}$ \cite{Li2012} and $B_{2g}$ symmetry \cite{Loret2019}. \textcolor{black}{Raman scattering probes large portions of the Brillouin zone, with distinctly different form factors in $B_{1g}$ and $B_{2g}$ symmetry.} In $B_{1g}$ symmetry, a broad feature with characteristic energy of $\sim 200$ meV ($\sim 1700$ cm$^{-1}$) was identified for a Hg1201 sample with a \textcolor{black}{slightly larger doping level ($p \approx 0.11$, $T_c = 77$ K) than in the present work} and associated with the two-magnon excitations of the undoped antiferromagnetic parent compounds. \textcolor{black}{We note that both paramagnon and charge excitations might contribute to the $B_{1g}$ response.} While a clear temperature dependence was observed below $\sim T^*$, this feature was seen to persist to higher temperatures \cite{Li2012}. 
In $B_{2g}$ symmetry, a feature of width $\sim 60 $ meV centered at $\sim 150$ meV ($\sim 1250$ cm$^{-1}$) was observed for Hg1201 \textcolor{black}{with $T_c = 72$ K, i.e., at nearly the same doping level as the present study,} and interpreted as a CDW energy scale \cite{Loret2019}. These values are remarkably close to \textcolor{black}{the intrinsic width of 77(16) meV (FWHM) and peak of 163(12) meV that we extract and associate with charge fluctuations (Fig. 3).
} 

\textcolor{black}{   
We now discuss the possibility that the dynamic charge and magnetic scales observed here correspond to features in the bosonic pairing glue function deduced from other spectroscopic techniques. While RIXS does not measure the glue function, it is a probe of both charge and magnetic excitations.
In an early optical conductivity study, it was suggested that deviations of the in-plane conductivity from the conventional Drude form may originate from coupling to a bosonic mode \cite{Thomas1988}.
Subsequent} optical spectroscopy efforts to extract the pairing glue revealed a robust peak around 50-60 meV and a second feature in the 100-300 meV range in a wide range of cuprates, including optimally-doped Hg1201 \cite{Dordevic2005,Hwang2007,vanHeumen2009,dalConte2012}. 
\textcolor{black}{Given that optical spectroscopy yields an average over the Brillouin zone, whereas our RIXS experiment focuses on the vicinity of $\boldsymbol{q}_\mathrm{CDW}$, this result is fully consistent with an interpretation of our data in terms of charge modes that contribute to the pairing glue. This is demonstrated in Fig. 6, which compares the glue function extracted at 100 K for Hg1201 ($T_c = 97$ K) with our RIXS result for the dynamic response at $T = T_c = 70$ K. As noted in Section III, the results in Fig. 3(a,b) are equally well captured by intrinsic heuristic Gaussian and log-normal functional forms, and the latter is used for the comparison.
Also shown in Fig. 6 is our result for the energy dependence of the strength of the paramagnon response at 70 K (obtained from the data in Fig. 4), as well as the momentum-integrated, local dynamic magnetic susceptibility obtained from neutron measurements near the antiferromagnetic wavevector for a sample with nearly the same doping level \cite{Chan2016a} (see also Fig. 4b). Assuming azimuthal symmetry, the former is proportional to the 2D momentum-integrated paramagnon strength. The neutron data are available in absolute units, and the RIXS paramagnon result is scaled to match the neutron data just below 100 meV.
The RIXS data for the charge response, in turn, are scaled up by a factor of 10 for ease of comparison and, as noted, because the fully-2D-integrated dynamic charge response may indeed be an order of magnitude larger than what we observe in the immediate vicinity of $\boldsymbol{q}_\mathrm{CDW}$ \cite{Supplemental,Arpaia2019}. We also note that up to about 20\% of the nominal paramagnon signal could be dynamic charge correlations  \cite{Supplemental}.
As seen from Fig. 6, there exists an intriguing connection between the charge and magnetic scales obtained from our RIXS experiment and those of the pairing glue function extracted via inversion of optical spectroscopy data for Hg1201 \cite{vanHeumen2009}. 
Qualitatively similar results have been obtained from Raman scattering, and Fig. 6 includes a result for Bi2212 \cite{Muschler2010}. As noted, these spectroscopies yield complex averages over the Brillouin zone and do not distinguish charge from magnetic response. RIXS has a key advantage over these and other techniques, as it is a rather direct, momentum- and energy-resolved measure of the charge and magnetic correlations. 
We also note that an analysis of Raman spectra for LSCO revealed a relatively narrow peak in the glue function below $\sim 100 $ meV and a broad hump up to $\sim 500 $ meV, seen in both B$_{2g}$ and $B_{1g}$ symmetry channels \cite{Fanfarillo2016}. It was concluded that, in B$_{1g}$ symmetry, spin modes contribute more strongly and at all frequencies, whereas charge modes are particularly prevalent at low/intermediate frequency in the $B_{2g}$ channel. Tunneling data for optimally-doped Bi2212 ($T_c = 95$ K), the lowest doping level investigated, indicate a strong local maximum at $\sim 40 $ meV and weaker, broad feature peaked just above 100 meV \cite{Ahmadi2011}.
}

\begin{figure}[t]
\hspace*{0mm}\includegraphics[width=8.5cm]{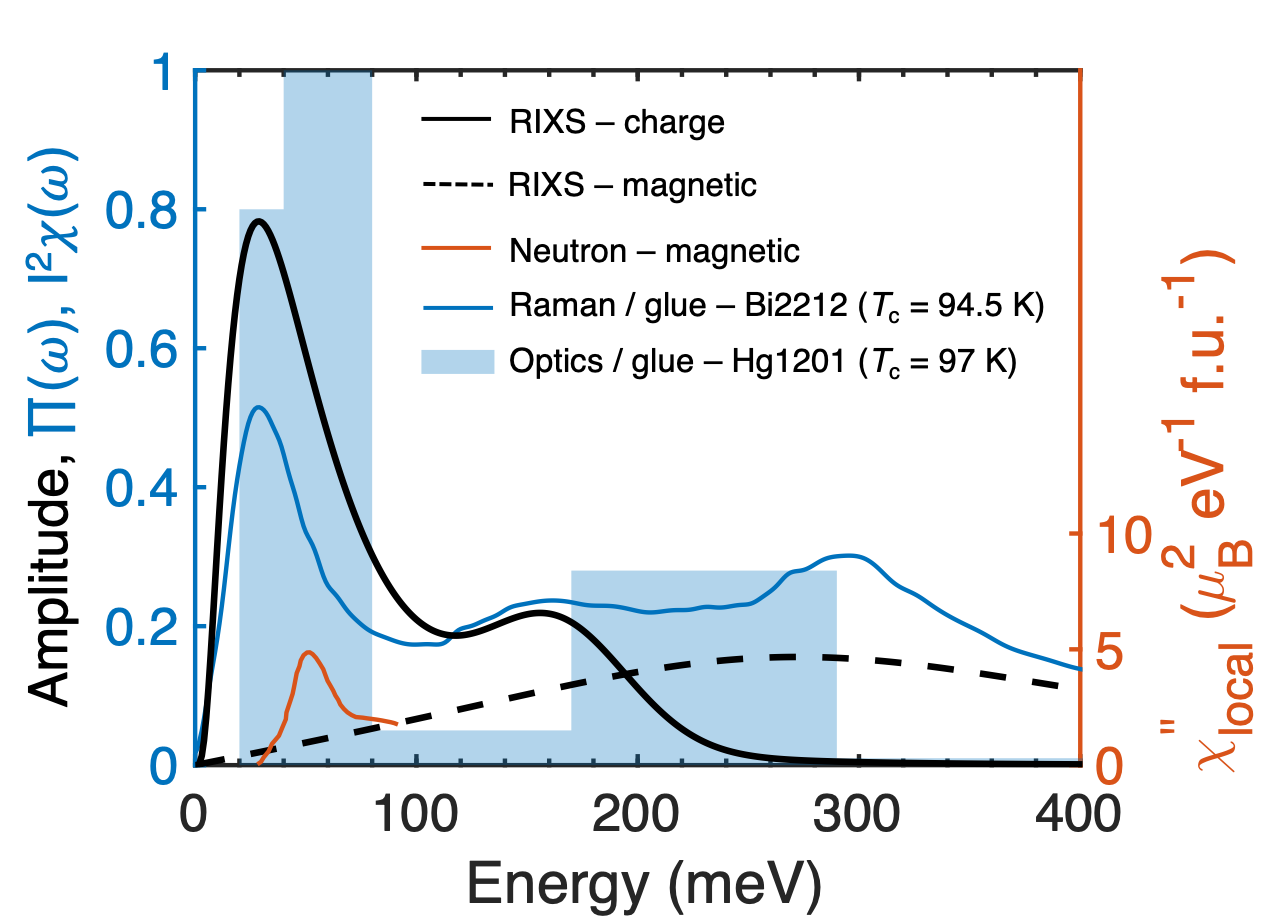}
\vspace{-3mm}
\caption{\textcolor{black}{
Comparison of RIXS results for dynamic charge and paramagnon response in Hg1201 with estimates for the superconducting pairing glue. The black solid and dashed lines are the amplitudes of the charge and magnetic excitations in Hg1201 at $T = T_c = 70$ K from the present work. As discussed in the text, the charge signal is multiplied by a factor of 10. The red solid line indicates the local magnetic susceptibility of Hg1201 measured by neutron scattering at the nearly same doping level \cite{Chan2016a}. The blue shaded area is the bosonic glue function extracted from optical  measurements of optimally-doped Hg1201 ($T_c$ = 97 K) at 100 K \cite{vanHeumen2009}; in this work, the glue function was not extracted for underdoped Hg1201, but it was established from analysis of other cuprates that the two characteristic energy scales are rather robust in the relevant doping range. The solid blue line is a glue function estimate for Bi2212 from Raman spectroscopy \cite{Muschler2010}, again at optimal doping; results in $B_{1g}$ and $B_{2g}$ symmetry are weighted 2:1 to best highlight the close correspondence of the energy scales with the RIXS result.    
}}
\vspace{0mm}
\label{Fig. 5}
\end{figure}

\textcolor{black}{Finally, we note that a recent RIXS study of electron-doped NCCO suggests a coupling between dynamic magnetic and charge correlations \cite{daSilvaNeto2018}.  In hole-doped cuprates, on the other hand, there is no clear evidence for such a coupling, except for the special case of the ``214" family of materials that exhibit charge-spin stripe order \cite{Miao2017}. 
While we can not rule out a small (20\% or less) charge contribution to the nominal paramagnon response for underdoped Hg1201 \cite{Supplemental}, we find no evidence of a significant coupling: the extracted paramagnon energy (Fig. 4) and damping coefficient  (Fig. S4 in \cite{Supplemental}) are insensitive to the dynamic charge correlations, consistent with the fact that the charge and magnetic excitations have somewhat different characteristic energy scales at/near $\boldsymbol{q}_{CDW}$.
}

\section{Conclusion}
In conclusion, we have measured the charge dynamics \textcolor{black}{and paramagnon response along the Cu-O bond direction} of underdoped Hg1201 using RIXS at the Cu $L_3$ edge with \textcolor{black}{high} energy resolution. This has allowed us to discern dynamic \textcolor{black}{charge correlations from the quasistatic CDW response}. Above $T_\mathrm{CDW}$, the temperature previously identified with the onset of CDW correlations, the response is purely dynamic, with an energy scale comparable to both the superconducting gap and the low-energy PG. As expected, quasistatic CDW correlations are observed at low temperature. However, there also exists an additional dynamic signature with a remarkably high energy scale that appears to be an imprint of the high-energy PG and associated with the strong electronic correlations that cause the charge-transfer gap of the undoped parent insulators.
\textcolor{black}{It is a distinct possibility that the dynamic charge correlations identified here have a different underlying physical mechanism than the CDW phenomenon and that they significantly contribute to the superconducting pairing glue}.
The present work sets the foundation for future RIXS measurements of the detailed temperature and doping dependence of the dynamic CDW correlations in the model cuprate Hg1201 \textcolor{black}{and, more broadly, for efforts to identify the pairing glue and} understand the CDW phenomenon of the cuprates in the context of large local charge fluctuations. 

\vspace{5mm}
\section{Acknowledgements}
We thank the ESRF for the allocation of beam time at ID32, D. Pelc for extensive discussions, and N. Bari\u{s}i\'{c}, A. V. Chubukov, G. Ghiringhelli, J. Haase, M.-H. Julien, Yuan Li, D. Pelc, D. Reznik, and E. H. da Silva Neto for valuable comments on the manuscript. The work at University of Minnesota was funded by the Department of Energy through the University of Minnesota Center for Quantum Materials, under DE-SC-0016371. The work at the TU Wien was supported by FWF project P27980-N36 and the European Research Council (ERC Consolidator Grant No 725521).

\bibliography{references}



\section*{Supplementary material}

## 1. Experimental Details

Resonant Inelastic X-ray Scattering (RIXS) measurements were performed at the ID32 beamline of the European Synchrotron Radiation Facility (ESRF) with energy resolution of 60 meV, as determined from the full-width at half-maximum (FWHM) of the non-resonant spectrum of a standard polycrystalline silver sample. The Hg1201 single crystal was cleaved *ex situ* prior to the RIXS measurements. The hole concentration was estimated from the $T_c$ vs $p$ relationship established in Ref. [S1]. Fig. S1(a) shows a schematic of the scattering geometry. Momentum scans were performed by rotating the sample about the **b** axis with the detector angle fixed at $2\theta = 150°$. We obtained energy-momentum intensity maps along $[H,0]$ for the $H$ values indicated in Fig. S1(b). At each momentum transfer, we measured the excitation spectrum for approximately 45 minutes. Additionally, X-ray absorption spectra were periodically taken to ensure that the incident X-ray energy remained at the maximum of the Cu $L_3$ resonance edge. Raw data are shown in Fig. S2.

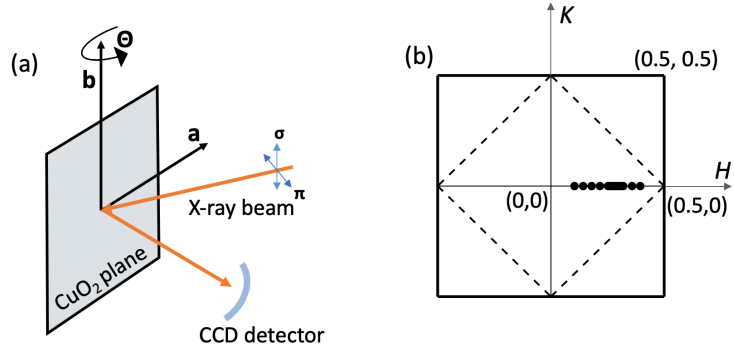

Fig. S1. (a) Scattering geometry with two different incident photon polarizations: parallel ($\pi$) and perpendicular ($\sigma$) to the $(H,0,L)$ scattering plane. (b) Reciprocal-space schematic with first nuclear and antiferromagnetic Brillouin zones (solid and dashed lines, respectively). Circles: $H$ values accessed in the experiment.

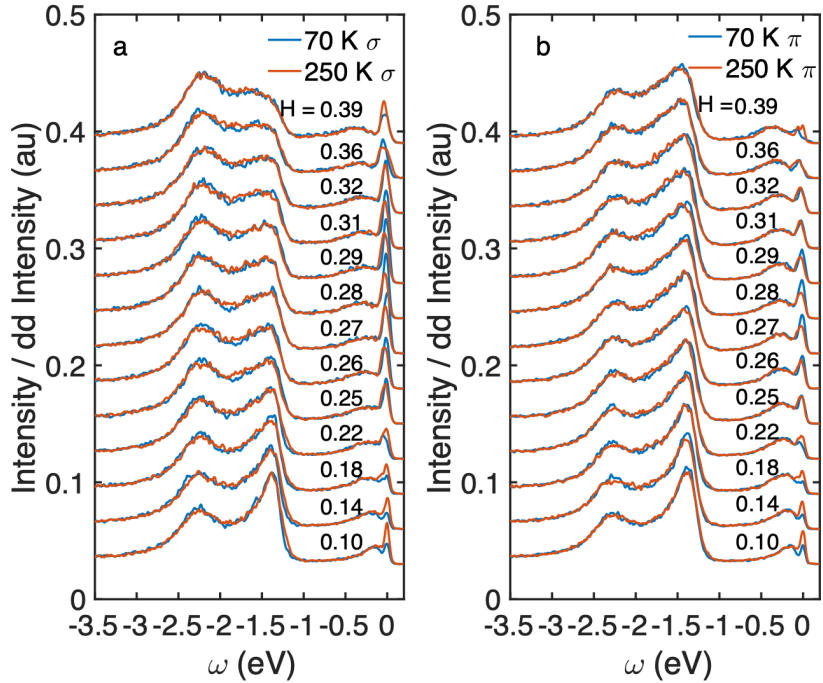

Fig. S2. RIXS spectra measured at 70 K and 250 K with (a) $\sigma$- and (b) $\pi$-polarized incident photons.

## 2. Paramagnons

In order to extract the paramagnon response, we decompose the mid-infrared region of the RIXS spectra into three components (Fig. S3) using the heuristic formula [S2, S3]:

$$\frac{I}{I_{dd}}(\omega) = G(\omega) + n_B L(\omega) + n_B \chi''(\omega),$$

where $n_B = [1 - \exp(-\hbar\omega/k_B T)]^{-1}$ is the Bose factor, $G(\omega)$ a resolution-limited Gaussian function that captures the elastic peak, $L(\omega)$ a resolution-limited Lorentzian profile used to effectively capture the phonon contribution, and the paramagnon was fitted to a damped harmonic oscillator form,

$$\chi''(\omega) = \chi_0'' \frac{\gamma\omega}{[\omega^2 - \omega_0^2]^2 + \omega^2 \gamma^2} = \frac{\chi_0''}{2\omega_1} \left[ \frac{\gamma/2}{(\omega - \omega_1)^2 + (\gamma/2)^2} - \frac{\gamma/2}{(\omega + \omega_1)^2 + (\gamma/2)^2} \right],$$

with damping coefficient $\frac{\gamma}{2} = \sqrt{\omega_0^2 - \omega_1^2}$. In addition, a linear background is assumed. In order to model the RIXS spectra, $G(\omega)$, $L(\omega)$ and $\chi''(\omega)$ are further convoluted with the instrument resolution function. The extracted damping coefficient shows no discernible momentum and temperature dependence (Fig. S4).

Figure S4 also compares the magnetic dispersion along [H,0] with results for doped and undoped La$_{2-x}$Sr$_x$CuO$_4$. The paramagnon dispersions for LSCO ($p = x = 0.11$) [S4] and Hg1201 ($p \approx 0.086$) agree within error up to at least $H = 0.35$ rlu.

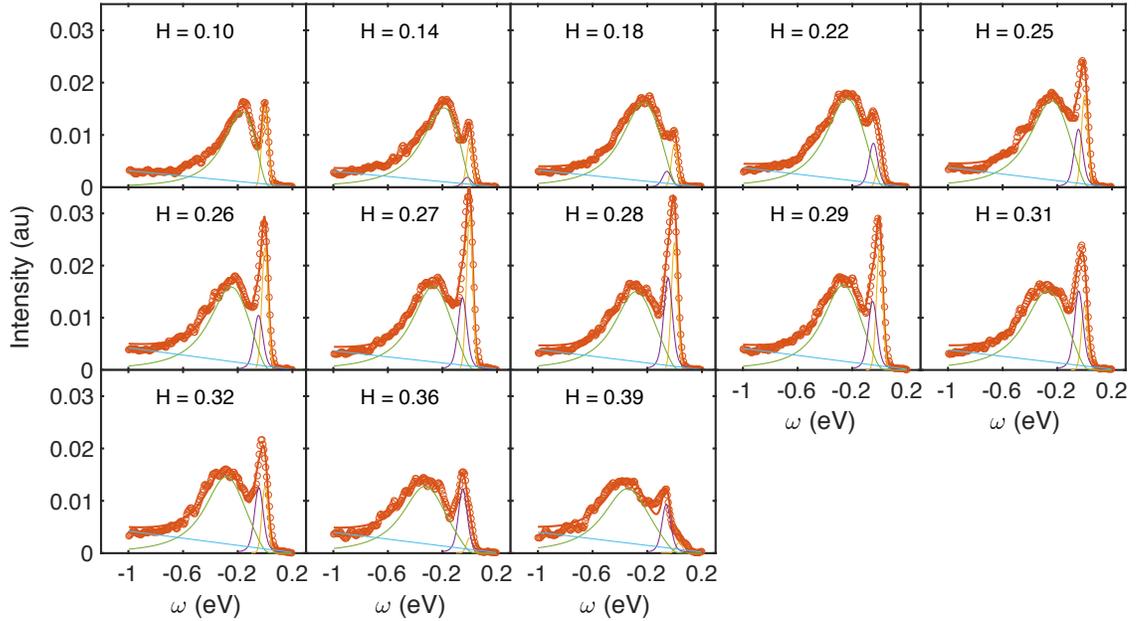

Fig. S3. Fits to RIXS data taken at 70 K in π geometry, in which the paramagnon response is strongest, with in-plane momentum transfer $H$ indicated. The spectra are decomposed into three components plus a linear background (blue): a resolution-limited elastic peak (yellow), a resolution-limited effective phonon peak (purple), and a damped paramagnon excitation (green).

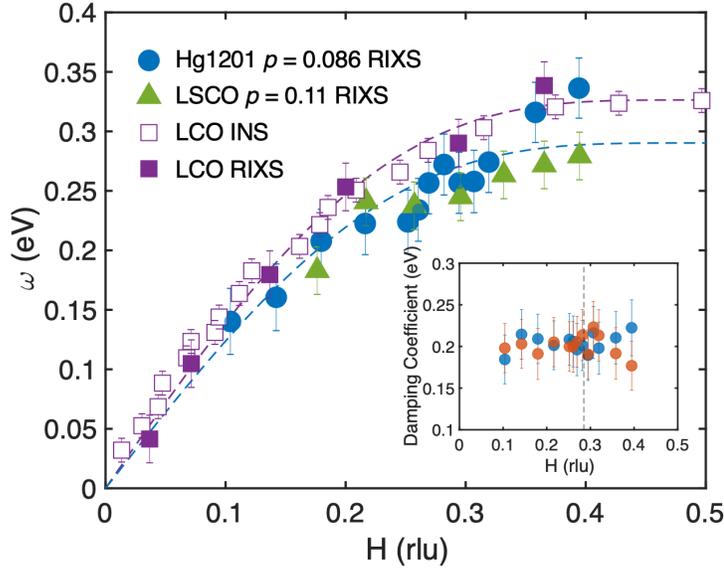

Fig. S4. Dispersion of magnetic excitations in Hg1201 ($p \approx 0.086$) at $T = 70$ K, compared with undoped antiferromagnetic $La_2CuO_4$ (LCO) [S2, S5, S6] and 11% Sr-doped $La_{2-x}Sr_xCuO_4$ (LSCO) [S4]. Blue and purple dashed lines are simple heuristic fits to nearest-neighbor spin-wave theory for Hg1201 and LCO, respectively. Inset: Damping coefficient for Hg1201 at 70 K and 250 K, with error bars set to the energy resolution (30 meV, half of the FWHM energy resolution). Neither the paramagnon dispersion nor the damping coefficient shows anomalous behavior at $\mathbf{q}_{CDW}$.

### 3. Polarization dependence

The Cu $L_3$-edge RIXS cross section can be calculated using the approximation of isolated $Cu^{2+}$ ions in a tetragonal crystal field with $D_{4h}$ symmetry. Following ref. [S7], we calculate the scattering amplitude for the present experimental geometry. For spin-conserving scattering,

$$F_{\sigma\sigma'} \propto \frac{1}{3}$$

$$F_{\pi\pi'} \propto -\frac{1}{3}\sin\left(\frac{tth}{2} - \delta\right)\sin\left(\frac{tth}{2} + \delta\right)$$

and for spin-flip scattering,

$$F_{\sigma\pi'} \propto -\frac{i}{6}\sin\left(\frac{tth}{2} - \delta\right)$$

$$F_{\pi\sigma'} \propto -\frac{i}{6}\sin\left(\frac{tth}{2} + \delta\right)$$

where $\sigma/\pi$ and $\sigma'/\pi'$ denote the incident and scattered photon polarization. The angles $tth$ and $\delta$ are the scattering angle and the rotation angle of the sample from the specular condition ($i.e., \delta = \theta - \frac{tth}{2}$), respectively. By comparing the intensity ratio of two different incident photon polarizations, $I_\sigma/I_\pi$, we can determine the charge or magnetic nature of the excitation. Figure S5 shows the theoretical and experimental polarization dependences in the quasielastic and paramagnon energy range. The comparison demonstrates the respective dominant charge and magnetic origin of the response. We note that this analysis points to the possibility that at both 70 K and 250 K up to 20% of the response in the paramagnon range is due to charge scattering.

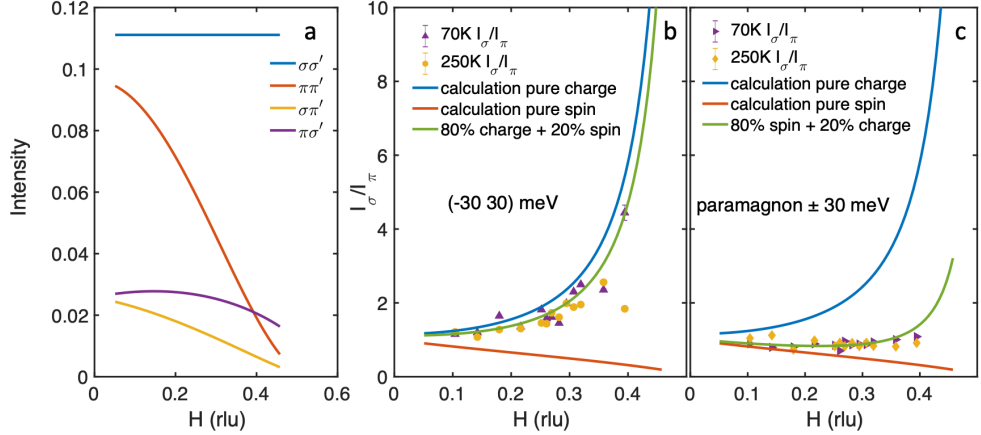

Fig. S5. (a) RIXS intensity calculation assuming isolated $Cu^{2+}$ ions in different scattering geometries. (b) Intensity ratio $I_\sigma/I_\pi$ in the quasielastic energy range. (c) Intensity ratio $I_\sigma/I_\pi$ in the paramagnon range. Yellow and purple circles: data. Solid lines: calculation.

## 4. Comparison with $Nd_2Ba_3CuO_{7-\delta}$

Figure S6 compares our result for Hg1201 (Fig. 3a) with data for $NdBa_2Cu_3O_{7-\delta}$ (NBCO) [S8]. In contrast to our data analysis for Hg1201 (Figs. 2 and 3), which removes featureless phonon contributions from the $H$-dependent intensity at/near $\mathbf{q}_{CDW}$, the NBCO data were analyzed under the assumption that phonon contributions at $\mathbf{q} = (H,0)$ can be removed by subtracting data at the same $|\mathbf{q}|$ along $\mathbf{q} = (H,H)$. Although the UD60 ($T_c = 60$ K) sample studied in ref. [S8] has a doping level that is closer to our Hg1201 ($T_c = 70$ K) sample, we consider here the phonon-subtracted NBCO data, which are only available for the optimally-doped sample NBCO OP90 ($T_c = 90$ K) [S8]. The data in Fig. S6 are scaled to match the low-temperature amplitudes. We observe that the relative strengths of the signal at $T_c$ and 250 K is very similar in both cases. Both compounds show quasielastic CDW correlations at $T_c$. The data analysis of ref. [S8] involved theoretical considerations of nearly critical CDW correlations. At 250 K and 150 K, the characteristic energy of dynamic charge correlations for NCCO OP90 was estimated to be 15 meV, whereas at $T_c$ a value of 7 meV was obtained [S8]. In Fig. S6, we show that the NBCO OP90 data at 250 K can also modeled as single-mode Stokes and anti-Stokes excitations with energy 18(3) meV, consistent with result in ref. [S8].

However, as noted, the analysis of NBCO [S8] involved the subtraction of RIXS spectra along [H,H] from the data of interest along [H,0]. We note that phonon scattering along [H,H] and [H,0] is not necessarily equivalent, that the required sample rotation might introduce systematic errors, and that elastic diffuse scattering from the sample surface may have an effective momentum and temperature dependence. In Fig. S7, we fit the phonon-subtracted RIXS spectra for NBCO OP90 to the sum of a Gaussian quasielastic peak (to effectively capture the quasistatic CDW response, a possible net low-energy phonon contribution, and spurious elastic diffuse scattering) and a Gaussian excitation profile with Stokes and anti-Stokes components (to capture a dynamic mode, convoluted with the Gaussian energy resolution). The data at 90 K, 150 K and 250 K are *simultaneously* fit, with the same excitation energy ($\omega_0$), width ($\gamma_0$), and amplitude ($A_0$, further corrected by the thermal Bose factor). In case that the phonon subtraction is incomplete, or of a possible coupling between low-energy phonons and the quasistatic CDW correlations, we allow the width of quasielastic peak ($\gamma_{el}$) to be slightly larger than the energy resolution (40 meV, FWHM). We allow the amplitudes of the elastic peaks ($A_{el}$) to vary at different temperatures to capture the onset of quasistatic CDW correlations. Given the sources of systematic errors noted above, the fits give a good description of the NBCO OP90 data, as seen from Fig. S7. This analysis yields a dynamic charge excitation scale of $\omega_0 = 38(4)$ meV,

consistent with our result for the lower of the two dynamic energy scales in Hg1201, and it points to an onset of quasistatic CDW correlations between 150 K and 250 K in NBCO OP90.

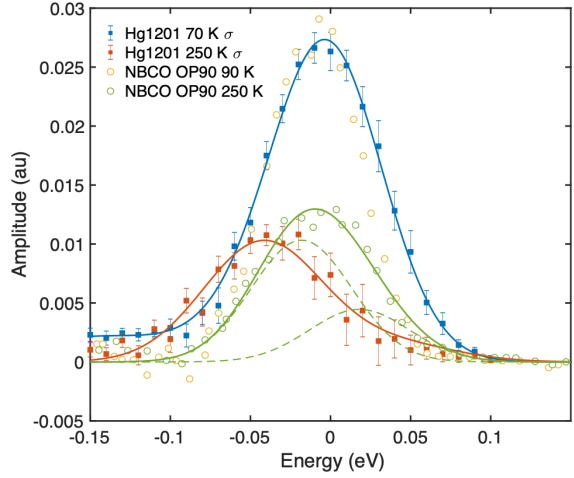

Fig. S6. Comparison of the scattering amplitude for optimally-doped NBCO OP90 ($T_c$ = 90 K) at **q** = (0.31,0), with data obtained at **q** = (0.22,0.22) subtracted [S8] (yellow and green circles), with the present result (Fig. 3a) for Hg1201 ($T_c$ = 70 K; blue and red squares). The solid green line is a heuristic fit to the NBCO OP90 250 K data assuming single-mode Stokes and anti-Stokes contributions (dashed lines). The peak is centered at 18(3) meV, consistent with the analysis result of 15 meV in ref. [S8].

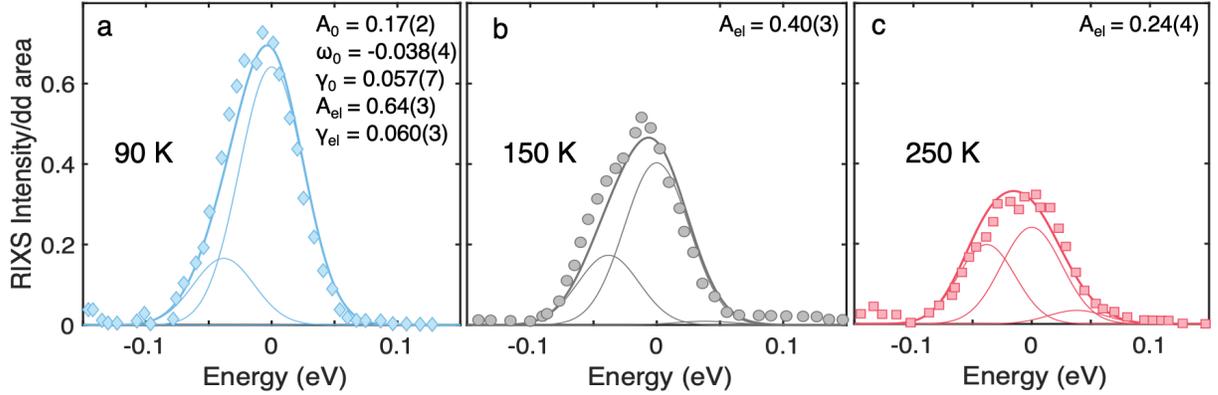

Fig. S7. Analysis of phonon-subtracted NBCO OP90 RIXS data as described in the supplementary text. The solid lines are fits to an effective Gaussian quasielastic peak plus an effective Gaussian excitation profile with Stokes and anti-Stokes components. As seen from the figure, the latter is only relevant at 250 K. No resolution-deconvolution is attempted. Except for the amplitude of the effective quasielastic peak ($A_{el}$), all parameters are globally fit, i.e., they are the same at all three temperatures. $A_0$, $\omega_0$ and $\gamma_0$ are the amplitude (corrected by the thermal Bose factor), characteristic energy, and FWHM of the dynamic charge mode; $\gamma_{el}$ is the FWHM of the effective quasielastic peak. The data are reasonably well described with a temperature-independent (up to the thermal Bose factor) dynamic charge response at all temperatures. This analysis yields $\omega_0$ = 38(4) meV and an intrinsic width of 41(10) meV, consistent with the present work for Hg1201.

## 5. Fit to Log-normal distribution

The analysis in Fig. 3 assumed a single mode at 250 K around 40 meV. The sum of Stokes and anti-Stokes contributions fits well to the data. As noted in the main text, the dynamic charge response can be equally well modeled as a distribution of modes of different energies. We show such an analysis in

Fig. S8, where a Log-normal distribution is assumed at zero temperature. Unlike the fit to the heuristic Gaussian form in Fig. 3, which yielded a large intrinsic width, the log-normal distribution is zero in the ω = 0 limit. The high-temperature (250 K) data can be modeled using Bose-Einstein statistics and detailed balance. The lineshape is further convoluted with the Gaussian resolution function. The log-normal distribution extracted from this analysis is used in Fig. 6.

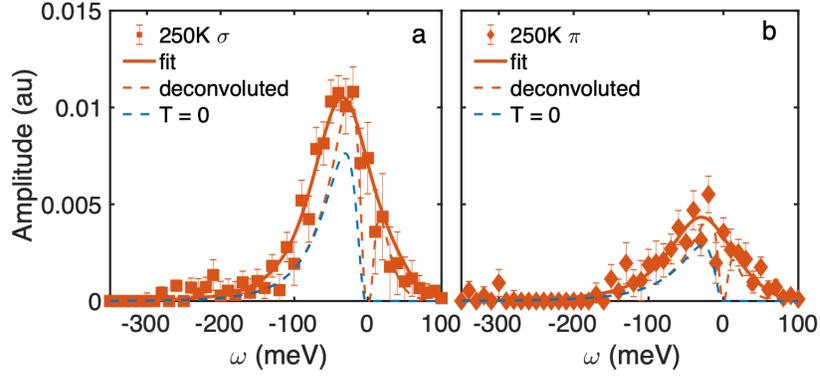

Fig. S8. Hg1201 250 K charge response modeled by a log-normal distribution in (a) σ- and (b) π-polarization. The experimental result (filled square and diamond symbols) is the same as in Fig. 3(a, b). The solid red lines are the results of fits to a resolution-convoluted log-normal distribution with a mean energy of 49(2) meV. The dashed red lines indicate the deconvoluted charge response at 250 K. The blue dashes line is the Bose-corrected charge susceptibility at zero temperature.

6. **EI-RIXS**

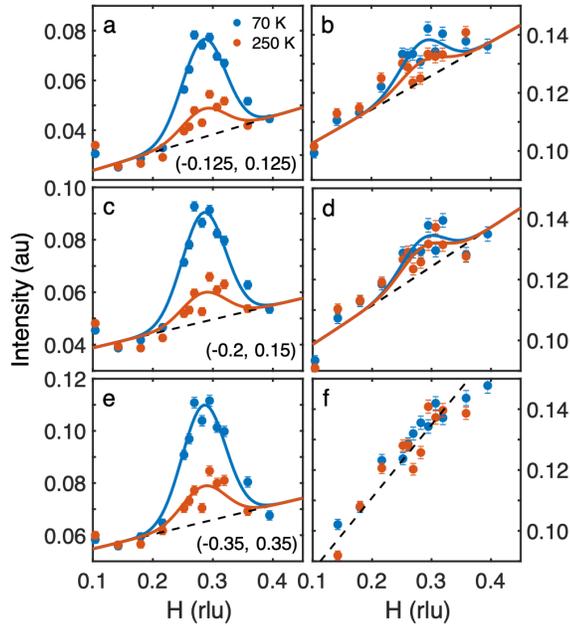

Fig. S9. RIXS spectra integrated over different energy windows. The left panels are the results of integration over the ranges (a) (-0.125, 0.125) eV, (c) (-0.2, 0.15) eV, and (e) (-0.35, 0.35) eV. The right panels are the corresponding complementary results of integration up to ± 1 eV with the respective ranges on the left excluded: (b) (-1, -0.125) & (0.125, 1) eV, (d) (-1, -0.2) & (0.15, 1) eV, and (f) (-1, -0.35) & (0.35, 1) eV. Solid red and blue lines are fits to a Gaussian peaks plus an underlying linear $H$ dependence (black dashed line).

Figure S9 shows RIXS data integrated over various energy ranges. The (-0.2,0.15) eV result corresponds to the range chosen in ref. [S8] and reveals a missing high-energy contribution (Fig. S9c,d). Charge correlations are relevant below ~350 meV (Fig. S9e,f,), consistent with Figs. 2 and 3.

## 7. High temperature CDW fluctuation in YBa$_2$Cu$_3$O$_{6+\delta}$ (YBCO)

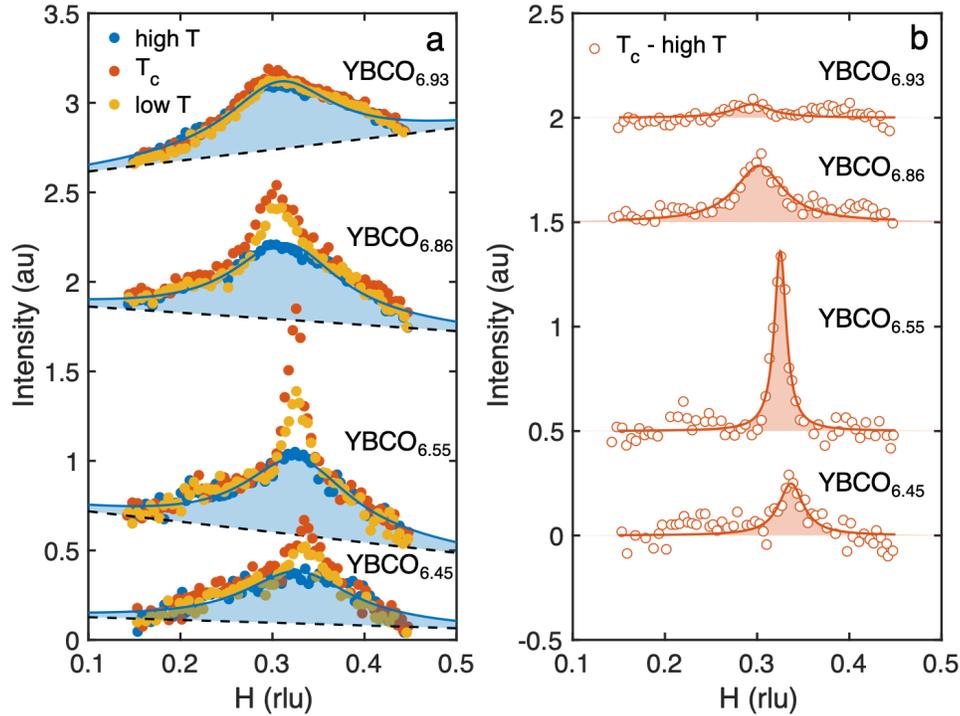

Fig. S10. Evidence for high-temperature charge fluctuations in YBCO measured by RXS. (a) Raw RXS spectra reproduced from [S9]. The broad peaks at high temperature are fit with a Lorentzian profile (solid blue line) on top of linear background (black dashed line). The shaded blue areas indicate the 1D spectral weight of the broad peak. (b) Spectra measured at $T_c$ after subtracting the broad peak in (a). The solid red lines are fits to a narrow Lorentzian peak, associated with quasistatic CDW correlations, and the red shaded areas indicate the 1D spectral weight.

Similar to the data for Hg1201, Bi2201 and NCCO summarized in Fig. 5, the extensively studied cuprate YBCO exhibits a local maximum around $q_{CDW}$ at high temperature. In Fig. S10, we show that the high-temperature YBCO RXS spectra of ref. [S9] are well described by a Lorentzian line shape. The nearly-incoherent high-temperature response is approximately independent of doping in the studied hole concentration range. Long-range CDW order is seen to grow on top of this nearly-incoherent charge scattering at low temperature, and in a relatively narrow doping range. The intensity of the incoherent charge scattering can then be calculated from its 2D spectral weight integrated over the whole Brillouin zone. For YBCO$_{6.55}$ and YBCO$_{6.86}$, the ratio of the 2D spectral weight at high temperature to that at low temperature is about 12, consistent with the recent results for NBCO in ref. [S8]. As discussed in the main text, the large underlying signal is dominated by dynamic charge correlations. One possibility why this signal is not seen in Fig. 5 might be that it is largely incoherent in single-layer cuprates.



\end{document}